\documentclass[12pt]{article}
    \usepackage{a4}
\usepackage[centertags]{amsmath}
\usepackage{amssymb}
\usepackage{amsthm}
\usepackage{bm}

\def\beq{\begin{equation}}
\def\eeq{\end{equation}}
\def\beqz{\begin{equation*}}
\def\eeqz{\end{equation*}}
\def\bea{\begin{eqnarray}}
\def\eea{\end{eqnarray}}
\def\nn{\nonumber}
\def\tr{ {\mbox{Tr} }}
\def\str{ {\mbox{Str} }}
\def\uu{ {\underline{{\mathbf{1}}}} }
\def\dd{ {\underline{{\mathbf{2}}}} }
\def\ttr{ {\underline{{\mathbf{3}}}} }
\def\dss{ {\delta_{\sigma\sigma'}} }
\def\pdss{ {\partial_\sigma\delta_{\sigma\sigma'}} }
\def\Ca{{C_{\uu\dd}}}

\def\C{{\cal C}}
\def\t{{\cal D}}
\def\tb{{\bar{\cal{D}}}}
\def\pb{{\bar{\partial}}}
\def\Ab{{\bar{A}}}

\def\as{{$AdS_5\times S^5\,$}}
\def\LL{{\cal L}}
\def\half{\frac{1}{2}}
\def\et{\qquad\mbox{and}\qquad}

\numberwithin{equation}{section}

\begin{document}
\begin{flushright}
AEI-2008-085
\end{flushright}

\vskip 0.5truecm

\begin{center}
{\large\bf The Classical Exchange Algebra of \as String Theory}

\vskip 0.5truecm

{\large{Marc Magro}}\\
\vskip 0.5cm
Universit\'e de Lyon, Laboratoire de Physique, ENS Lyon et CNRS UMR 5672,\\
46 all\'ee d'Italie, F-69364 Lyon CEDEX 07, France \\

\medskip

and

\medskip

Max-Planck-Institut f\"ur Gravitationsphysik\\%
Albert-Einstein-Institut\\%
Am M\"uhlenberg 1, 14476 Potsdam, Germany

\smallskip

{\tt Marc.Magro@ens-lyon.fr}\\
\end{center}

\begin{abstract}
\noindent The classical exchange algebra satisfied by the monodromy matrix of $AdS_5\times S^5$ string theory in the Green-Schwarz formulation is determined by using a first-order Hamiltonian formulation and by adding to the Bena-Polchinski-Roiban Lax connection terms proportional to  constraints. This enables in particular to show that the conserved charges of this theory are in involution. This result is obtained for a general world-sheet metric. The same exchange algebra is obtained within the pure spinor description of \as string theory. These results are compared to the one obtained by A. Mikhailov and S. Sch\"afer-Nameki for the pure spinor formulation.
\end{abstract}

\section{Introduction}

Integrability plays a key role in the understanding of the correspondence between string theory on $AdS_5\times S^5$ and superconformal $N=4$ Yang-Mills theory. For the AdS side of this correspondence, it has been proved in \cite{Bena:2003wd} that the classical equations of motion can be cast into a zero curvature equation satisfied by a Lax connection. This property leads to the existence of an infinite number of conserved charges. However, determining the Poisson brackets of these conserved charges has been a long-standing problem. As for any other integrable system, it is natural to expect that these charges are in involution, i.e. that their Poisson brackets (P.B.) vanish. Actually, from some conventional  point of view, it is a necessary condition in order to properly call this theory  an integrable one. For instance, for  finite dimensional systems, it is  a necessary condition  in order to apply Liouville's theorem (see for instance \cite{Babelon}). Note however that  at the quantum level, the commutation of the conserved charges is not necessary for the factorization of the S-matrix \cite{Parke:1980ki}. From the point of view of the AdS/CFT correspondence, it is very unlikely that the conserved charges would not be in involution. An early sign of this expected involution property is the observation that the dilatation operator of $N=4$ Yang-Mills theory corresponds to the Hamiltonian of an integrable quantum spin chain at first order in perturbation theory in the planar limit. This was first discovered in \cite{Lipatov:1993qn,Braun:1998id} and more specifically in the $N=4$ context, extending the discovery of \cite{Min-zar}, in 
\cite{Beisert:2003jj}. Evidence was then presented in \cite{Beisert:2003tq} that this integrability is present at higher orders.

It is therefore quite frustrating that this expected property of involution of the conserved charges has not yet been directly proved. Furthermore, it should in fact be almost as easy to prove that property as it has been to determine the Lax pair in \cite{Bena:2003wd}. Indeed, this should be a fundamental property of that theory. In other words, this involution property should neither be specific to the Green-Schwarz or to the pure spinor formulations nor be related to a specific gauge choice like, for instance, the conformal gauge but should be valid in full generality. We   show in this article that it is indeed the case. As it has been achieved for a subsector of \as,  the determination of the exchange algebra is also a necessary step towards the computation of action-angle variables \cite{Dorey:2006mx,Vicedo:2008jk} and  semi-classical quantisation \cite{Vicedo:2008jy,Vicedo:2008jk}.

\medskip

There has been many attempts to compute the classical exchange algebra and to prove the involution property. The most successful one has been developed within the pure spinor formulation in \cite{Mikhailov:2007eg}. We will explain in section \ref{sec33} why our result is different from the one obtained in \cite{Mikhailov:2007eg}. Furthermore, our approach is more direct and is valid for both Green-Schwarz and pure spinor formulations.  The  other attempts can be found in  \cite{Das:2004hy,Das:2005hp,Mikhailov:2006uc,Kluson:2007ua,Aoyama:2007tz}.  In section \ref{sec33}, the result obtained by A.~K.~Das et al. in \cite{Das:2004hy} is discussed relatively to our result. We will also argue at the end of \S\ref{sec24} that the approach chosen in \cite{Das:2004hy} to determine  the phase space variables appears in fact to be incomplete, contrary to the first-order formulation considered in the present work.

\medskip

The technical tool used in this article is a first-order Hamiltonian formulation of coset models. In this formulation, the dynamical variables are the currents instead of the group element. It is motivated at the beginning of section \ref{sec21} and presented in detail in section \ref{sec2}.  Let us however discuss immediately the main, and rather simple, idea used for the computation of the exchange algebra. For that, we recall basic properties of Lax connections. Consider a classical system whose Lagrangian equations of motion can be cast in the form of a zero-curvature equation
\beq
\partial_\alpha \LL_\beta -\partial_\beta \LL_\alpha -[\LL_\alpha,\LL_\beta]=0 \label{dim1}
\eeq
for a Lax pair $\LL_\alpha(\sigma,\tau;z)$ taking values in some Lie algebra. Here $(\sigma,\tau)$ are worldsheet coordinates and $z$ is a spectral parameter.
We recall that the monodromy matrix $T(\tau;z)$ is the path-ordered exponential of the spatial Lax component ${\cal L}_\sigma(\sigma,\tau;z)$:
\beq
T(\tau;z) = P \overleftarrow{\exp} \int_0^{2\pi} d\sigma {\cal L}_\sigma(\sigma,\tau;z) \label{defmon}
\eeq
where we consider periodic boundary conditions. It follows from the zero curvature equation (\ref{dim1}) and from the periodicity of the Lax connection in the variable $\sigma$ that
\beqz
\partial_\tau T(\tau;z) = [\LL_\tau(0,\tau;z),T(\tau;z)].
\eeqz
This implies that the eigenvalues of the monodromy matrix are integrals of motion. These conserved quantities are in involution only if the Poisson bracket $\{T(\tau;z),T(\tau,z')\}$ has a special form.  The classical exchange algebra corresponds then to this Poisson bracket. To determine it, one first needs to compute the P.B. of ${\cal L}_\sigma(\sigma,\tau;z)$ with ${\cal L}_\sigma(\sigma',\tau;z')$.

The zero curvature condition (\ref{dim1}) is invariant under the gauge transformations
\beq
\LL_\alpha \to \LL^U_\alpha  =U\LL_\alpha U^{-1} + \partial_\alpha U U^{-1}. \label{formdimtard}
 \eeq
 We call this invariance a formal gauge invariance to avoid confusion with the other gauge invariances present in \as String theory.  Under these transformations, $T$ transforms as $ T(\tau,z) \to U(2\pi,\tau) T(\tau,z) U^{-1}(0,\tau)$ and thus its eigenvalues are invariant.

\medskip

Consider now a Lagrangian system whose Legendre transformation leads to Hamiltonian constraints. This property holds in the case of \as String theory. Then, going from the Lagrangian to the Hamiltonian formulations, there is nothing that forbids  to add to the "Lagrangian" Lax pair terms proportional to the constraints.  Having in mind the general theory of constrained systems, this is actually an expected property. One can  give an argument in favor of this process. Indeed, from the Lax pair, we construct successively the monodromy matrix and the conserved quantities. But the Hamiltonian itself is a specific conserved quantity. And, as usual with constrained systems, it contains terms proportional to the constraints and the Lagrange multipliers.

One can in fact construct an infinite number of "Hamiltonian" Lax pairs. Of course, by definition, all these Lax pairs have the same value on the constraint surface. But, as usual with constrained systems\footnote{See \cite{Henneaux} for a general reference on constrained systems. The definitions of first and second class constraints and of the Dirac bracket are recalled in the appendix.}, their Poisson brackets will not be the same as one shall first compute the P.B. and only afterwards evaluate them  on the constraint surface. This discussion might sound rather strange to the reader as we are presently claiming that there are many different exchange algebras, in complete opposition with the title of this article ! There is however no contradiction. Once all unphysical degrees of freedom are eliminated, i.e. once we introduce all the necessary gauge fixing conditions to have a complete  system of second-class constraints in order to define Dirac brackets, all these Lax pairs will have the same Dirac bracket. This is so because by definition the Dirac brackets of the constraints with any phase space variable strongly vanish.  This is indeed only in the sense of Dirac bracket that one can talk about an unique exchange algebra.
 However, and this is the crucial point, there might be a better starting point to compute this algebra than just the one consisting in taking straightforwardly the Lax pair obtained   from the Lagrangian formulation. In other words, there might exist a  "natural" Hamiltonian Lax pair, whose Poisson brackets have  the simplest form. This is indeed what happens for \as String theory as   it will be shown in section \ref{sec3}.

It is also \label{pagecomment}necessary to make here another comment related to $\kappa$-symmetry. Under such a transformation, the Bena-Polchinski-Roiban Lax connection transforms by a formal gauge transformation (\ref{formdimtard}). This property is explicitly established in \cite{Arutyunov} for the $AdS_4 \times CP^3$ case (see also \cite{Berkovits:2004jw}). This means that the action (in the sense of P.B.) of a first-class constraint on a Lax connection should correspond to a particular case of a formal gauge transformation. It is therefore {\em a priori} expected that if $\LL$ and $\widetilde{\LL}$ are two Lax connections differing only through a term proportional to a first-class constraint, then the P.B. of $\LL$ should have the same form as the ones of $\widetilde{\LL}$. However, the term that will be added to the Bena-Polchinski-Roiban Lax connection is a mixture of first-class and second-class constraints. We will discuss more precisely this statement in section \ref{sec33}.

 \medskip

Let us now quickly review known forms of  P.B. of ${\cal L}$ that lead to involution of conserved charges and indicate which one we will find.
In the following, we simply  denote ${\cal L}(\sigma,z) \equiv {\cal L}_\sigma(\sigma,\tau;z)$.
The standard simple form of Poisson brackets\footnote{As these are equal-time P.B., the time dependence will not be indicated in this article.}   of ${\cal L}$ ensuring involution of the conserved charges  is (for a review, see \cite{Faddeev})
\beq
\{ {\cal L}_\uu(\sigma,z_1), {\cal L}_\dd(\sigma',z_2)\} = [r_{\uu\dd}(z_1,z_2), {\cal L}_+]\dss \label{sam123}
\eeq
where $\dss=\delta(\sigma-\sigma')$. We use  conventional tensorial notations $\LL_\uu = \LL \otimes 1$ and $\LL_\dd = 1 \otimes \LL$ (see section \ref{appa1})  and have introduced  $$\LL_\pm = \LL_\uu(\sigma,z_1) \pm \LL_\dd(\sigma,z_2).$$ For simplicity, we have considered a non-dynamical $r$-matrix i.e. which does not depend on the phase space variables but only on the spectral parameters.  The P.B. of $T$ are then:
$$
\{ T_\uu(z_1), T_\dd(z_2) \} = [r_{\uu\dd}, T_\uu(z_1)T_\dd(z_2)].
$$
 This implies that the traces $\tr[T^n(z_1)]$ and $\tr[T^m(z_2)]$ are in involution. For completeness, we recall that the $r$-matrix is antisymmetric\footnote{More precisely, $P r_{\uu\dd}(z_2,z_1) P = -r_{\uu\dd}(z_1,z_2)$ where $P(A\otimes B)P=B\otimes A$ for any matrices $A$ and $B$.} and that the Jacobi identity is satisfied when $r$ is a solution of the classical Yang-Baxter equation,
\beq
[r_{\uu\dd},r_{\uu{\underline{{\mathbf{3}}}}}]+[r_{\uu{\underline{{\mathbf{2}}}}},
r_{\dd{\underline{{\mathbf{3}}}}}]+[r_{\uu{\underline{{\mathbf{3}}}}},
r_{\dd{\underline{{\mathbf{3}}}}}]=0,\label{yb}
\eeq
where we have not explicitly indicated the spectral dependence. It is however clear that such a form does not hold for \as String theory.

A generalization of the  P.B. (\ref{sam123}) has been given by J.~M.~Maillet in \cite{Maillet:1985fn,Mail2,Maillet:1985ek} and is:
\begin{multline}
\{ {\cal L}_\uu(\sigma,z_1), {\cal L}_\dd(\sigma',z_2)\} = [r_{\uu\dd}(z_1,z_2), \LL_+]\dss -[s_{\uu\dd}(z_1,z_2),\LL_-] \dss \\
-2 s_{\uu\dd}(z_1,z_2) \pdss. \label{mail}
\end{multline}
Again,  we restrict ourselves to non-dynamical $r$ and $s$ matrices. We will call in the future this form of P.B. the $r/s$ form. $r$ is antisymmetric while $s$ is symmetric. A sufficient condition for the Jacobi identity to be satisfied is that $r$ and $s$ are solutions of the extended Yang-Baxter equation:
\beq
[r_{\uu{\underline{{\mathbf{3}}}}} + s_{\uu{\underline{{\mathbf{3}}}}},
r_{\uu\dd}- s_{\uu\dd}] + [r_{\dd{\underline{{\mathbf{3}}}}} + s_{\dd{\underline{{\mathbf{3}}}}}, r_{\uu\dd}+ s_{\uu\dd}] +[r_{\dd{\underline{{\mathbf{3}}}}}+s_{\dd{\underline{{\mathbf{3}}}}}, r_{\uu{\underline{{\mathbf{3}}}}}+s_{\uu{\underline{{\mathbf{3}}}}}]=0. \label{YB2}
\eeq
Contrary to the P.B. (\ref{sam123}), the P.B. (\ref{mail}) involves non-ultra-local terms. As a consequence, the P.B. of the monodromy matrix are not well defined. This is the famous problem related to non-ultra-local terms. It is possible to regularize\footnote{This regularization is however not completely satisfactory as the Jacobi identity (for P.B. involving the monodromy matrix)
 is not fully satisfied. It is only "weakly" satisfied (see \cite{Maillet:1985ek} for details). This regularization has been however  successfully used in \cite{Dorey:2006mx}.} the P.B. of the monodromy matrix  \cite{Maillet:1985ek}. In that case one gets
\beqz
\{ T_\uu(z_1), T_\dd(z_2) \} = [r_{\uu\dd}, T_1(z_1)T_2(z_2)] + T_\uu(z_1) s_{\uu\dd} T_\dd(z_2) - T_\dd(z_2) s_{\uu\dd} T_\uu(z_1),
\eeqz
which again leads to the involution of $\tr[T^n(z)]$. Note that the vanishing of the P.B. of $\tr[T^n(z_1)]$ with $\tr[T^m(z_2)]$ is independent of the regularization chosen \cite{Maillet:1985ek}.

We will show that the  P.B. of the Hamiltonian Lax spatial component of  \as String theory has the $r/s$ form.

\medskip

The plan of this paper is the following. In section \ref{sec2}, we start by discussing the first-order Hamiltonian formulation for the principal chiral model. The goal is to present this formulation for the simplest case and to show how  the  P.B. of the currents of this model are recovered. This method is then applied for pedagogical reasons to a bosonic coset $G/H$ model. Indeed, we will discuss there a property related to the gauge symmetry of this model. This property has a more complicated analogue (related to $\kappa$-symmetry) in the Green-Schwarz formulation of \as String theory. The first-order Hamiltonian  technique is applied in section \ref{sec23} to the pure spinor case. The corresponding analysis for the Green-Schwarz case is presented in the next section. Note that   sections  \ref{sec23} and \ref{sec24} can be read independently. For the Green-Schwarz formulation, we start with a  general world-sheet metric and a general coefficient, $\kappa$, in front of the Wess-Zumino term present in the Lagrangian of this theory. Making the first-order analysis, we obtain primary constraints and, in order to ensure  stability of these constraints, a secondary one. We show then that, when $\kappa =\pm1$, i.e. when $\kappa$-symmetry is present, there is no further constraint. In the next step, we partially gauge-fix the theory and eliminate variables that are redundant within the first-order Hamiltonian formulation. The results of sections \ref{sec23} and \ref{sec24} include the canonical variables, the constraints they satisfy, and their Hamiltonians.

In section \ref{sec3}, we start by introducing the Hamiltonian Lax connection
 and compute in \S\ref{sec32bis} the Poisson brackets of its spatial component. The main results of this article correspond to the equations (\ref{finalmbis}) and (\ref{theresult})-(\ref{bepulsebis}). In \S\ref{sec33},  these  results are compared to the ones obtained in \cite{Mikhailov:2007eg} and in \cite{Das:2004hy}. Finally, we make some comment on the link between the Green-Schwarz and pure spinor formulations.

The appendix contains definitions and technical results used in sections \ref{sec2} and \ref{sec3} and a reminder on constrained systems.

\section{First-Order Formulation} \label{sec2}
\subsection{Principal Chiral Model}\label{sec21}

The Lagrangian of the principal chiral model (PCM) is
\beqz
L = {1\over 2} \bigl(g^{-1} \partial_0 gg^{-1} \partial_0 g -g^{-1} \partial_1 gg^{-1} \partial_1 g\bigr)
\eeqz
where $g(\sigma,\tau)$ takes value in some semi-simple Lie group, and where taking the trace over the corresponding Lie algebra is understood. The equations of motion are then $\partial_0 (g^{-1} \partial_0 g) - \partial_1 (g^{-1} \partial_1 g)=0$.

\paragraph{Motivation}
  We \label{motupage} are only interested in determining the P.B. of
the currents $A_\alpha = - g^{-1} \partial_\alpha g$. Indeed, the Lax connection depends on $g$ only through $A_\alpha$. It is therefore desirable to compute directly these P.B. without having to introduce coordinates on the Lie group. One approach to do so, and which is used for instance in \cite{Das:2004hy,Bianchi:2006im}, is the following. Consider $A_1$ as the only dynamical variable. Rewrite formally the Maurer-Cartan equation, $\partial_0 A_1 - \nabla_1 A_0 =0$, satisfied by the currents as $A_0 = \nabla_1^{-1}(\partial_0 A_1)$. We have introduced here the covariant derivative $\nabla_1 = \partial_1 - [A_1,]$. Compute then the conjugate momentum of $A_1$. However,  we will explain in \S\ref{sec33} why this procedure can be considered as incomplete when constraints are present:
 It gives the right P.B. for part of the currents  but does not give any information for the remaining components. We will therefore proceed differently.

\paragraph{Lagrangian Equations}

The starting point is the Lagrangian equations satisfied by the currents. Those are  the Maurer-Cartan equation,
\beq
\partial_0 A_1 - \nabla_1 A_0 =0, \label{zero9}
\eeq
and  the equation of motion
\beq
\partial_0 A_0 - \partial_1 A_1 =0. \label{eqmpcma}
\eeq
We start then with the Lagrangian\footnote{I thank N. Beisert for suggesting this approach.}
\beqz
L = {1\over 2} (A_0A_0-A_1A_1) + \Lambda(\partial_0 A_1 - \nabla_1 A_0)
\eeqz
where the independent dynamical variables are now $(A_0,A_1,\Lambda)$. It is clear that the equation of motion of the Lagrange multiplier $\Lambda$ implies\footnote{Up to some global problems not considered here.} $A_\alpha = -g^{-1}\partial_\alpha g$. Thus, at least classically, this theory is equivalent to the PCM.

\paragraph{Primary and Secondary Constraints}
Let us now do the Legendre transformation and the Hamiltonian analysis. One finds the constraints\footnote{The notation $\approx$ stands for "on the constraint surface".}
\beqz
\Pi_0 \approx 0, \qquad \Pi_1 -\Lambda \approx 0,\qquad \Pi_\Lambda \approx 0
\eeqz
with obvious notations. The Poisson brackets of the canonical variables are written in the appendix. The Hamiltonian density $h$ is then
\beq
h = -{1\over 2} (A_0A_0-A_1A_1) + \Lambda\nabla_1 A_0 + \alpha \Pi_0 + \beta(\Pi_1 - \Lambda) + \gamma \Pi_\Lambda  + \mu {\cal C} \label{hpcm1}
\eeq
where $\alpha$, $\beta$, $\gamma$ and $\mu$ are Lagrange multipliers. In eq.(\ref{hpcm1}), we have already taken into account the secondary constraint
\beq
{\cal C} = A_0 + \nabla_1 \Lambda \approx 0,  \label{constraintc}
\eeq
coming from imposing stability of the primary constraint $\Pi_0$ under time evolution. Requiring that the constraints are preserved by the dynamics does not lead to further constraints and fixes all the Lagrange multipliers:
\begin{align*}
\mu &\approx 0, &\qquad
\beta &\approx \nabla_1 A_0,\\
\gamma &\approx - A_1 - [\Lambda,A_0], &\qquad
\alpha &\approx  \partial_1 A_1.
\end{align*}

\paragraph{Hamiltonian Equations of Motion}

The equations of motion
for $A_0$ and $A_1$ corresponding to the Hamiltonian $H =\int d\sigma h$ are\footnote{With the convention $\{ A, \Pi\} = \delta$ (see appendix).}:
\beqz
{ dA_1 \over d\tau} = \beta \approx \nabla_1 A_0 \et  { dA_0 \over d\tau} = \alpha \approx \partial_1 A_1.
\eeqz
So they coincide respectively with the Lagrangian equations (\ref{zero9}) and  (\ref{eqmpcma}).

\paragraph{Elimination of Variables}

The first thing to do is to get rid of $\Lambda$ and its momentum conjugate. This is easy as the constraints $\Pi_1-\Lambda$ and $\Pi_\Lambda$ form a set of second-class constraints, and as it means that we can simply forget about $\Pi_\Lambda$ and replace everywhere $\Lambda$ by $\Pi_1$. We have thus now the canonical variables $(A_0,A_1,\Pi_0,\Pi_1)$ together with the two constraints
\beq
\Pi_0 \approx 0, \qquad {\cal C} =A_0 + \nabla_1 \Pi_1 \approx 0. \label{constrong}
\eeq
These constraints form a set of second-class constraints. Indeed, the matrix of their P.B.
\beq
\begin{array}{cccc}
&&\Pi_{0\dd}(\sigma')& {\cal C}_\dd(\sigma')\\
\hline
\Pi_{0\uu}(\sigma)&|& 0 & -C_{\uu\dd} \dss \\
{\cal C}_\uu(\sigma)&|&C_{\uu\dd} \dss&[C_{\uu\dd}, (\nabla_1 \Pi_1)_{\dd}] \dss
\end{array} \label{matdel}
\eeq
is invertible. This matrix is written in tensorial notation,  $C_{\uu\dd}$ being the quadratic Casimir (see appendix for further definitions). Therefore, we can put the constraints $(\Pi_0, {\cal C})$ strongly to zero and compute the corresponding Dirac brackets for the currents $(A_0,A_1)$. The definition of the Dirac bracket is recalled in eq.(\ref{defdiracapp}). Although it is quite instructive to make explicitly this computation, we will use in fact a shortcut.  Indeed, a better interpretation of putting the constraints (\ref{constrong}) strongly to zero, is that we are then left with the  variables $(A_1, \Pi_1)$ and that $A_0$ is now completely identified with $-\nabla_1 \Pi_1$. Furthermore, due to the form of the matrix (\ref{matdel}), the variables $(A_1, \Pi_1)$ have the same Poisson and Dirac brackets. This means that they remain canonical  with respect to the Dirac bracket. We then have:
\begin{align*}
\{ A_{1\uu}(\sigma), A_{1\dd}(\sigma') \} &= 0,\\
\{ (\nabla_1 \Pi_1)_{\uu}(\sigma), A_{1\dd}(\sigma') \} &= \bigl[C_{\uu\dd}, A_{1\dd} \bigr] \dss - C_{\uu\dd} \pdss,\\
\{ (\nabla_1 \Pi_1)_{\uu}(\sigma) ,  (\nabla_1 \Pi_1)_{\dd}(\sigma') \} &=
 \bigl[C_{\uu\dd},   (\nabla_1 \Pi_1)_{\dd} \bigr] \dss.
\end{align*}
We    recover  in that way the P.B. of the currents of the PCM. Finally, starting from the expression (\ref{hpcm1}), the Hamiltonian can be rewritten as:
\beqz
H = \int d\sigma \Bigl[-\half (A_0A_0 -A_1A_1)+ \Pi_1 \nabla_1 A_0\Bigr]
={1\over 2} \int d\sigma  \bigl(\nabla_1 \Pi_{1}\nabla_1 \Pi_{1}+A_1A_1\bigr).
\eeqz
For that, we have  used the constraints (\ref{constrong}) and integrated by parts. Thus, we do recover both the Hamiltonian and the P.B. of the PCM.

\subsection{Bosonic Coset $G/H$ Model}\label{sec22}

As this section is only here for pedagogical purpose, we skip unimportant details and concentrate on important points relevant for a better understanding of the  \as case. Let ${\cal G}$ be the Lie algebra associated with $G$. To make contact with the \as case, we denote by ${\cal H} ={\cal G}^{(0)}$ the Lie subalgebra corresponding to $H$ and write also the decomposition ${\cal G} = {\cal G}^{(0)} \oplus  {\cal G}^{(2)}$ of ${\cal G}$ as a vector space. In particular, $[{\cal G}^{(i)}, {\cal G}^{(j)}] \subset {\cal G}^{(i+j \,mod\, \mathbb{Z}_2)}$. For $M \in {\cal G}$, we write $M = M^{(0)} + M^{(2)}$.

\paragraph{Lagrangian Equations}
The equations satisfied by the currents are the Maurer-Cartan equation
\beq
\partial_0 A_1 = \nabla_1 A_0 \label{zerogh}
\eeq
and the equation of motion
\beq
\partial_0 A_0^{(2)} - \partial_1 A_1^{(2)} - [ A_0^{(0)},A_0^{(2)}] + [ A_1^{(0)}, A_1^{(2)}]=0.  \label{lagbcm00}
\eeq
The starting point of our analysis is the Lagrangian
\beq
L = {1\over 2} (A_0^{(2)}A_0^{(2)}-A_1^{(2)}A_1^{(2)}) + \Lambda(\partial_0 A_1 - \nabla_1 A_0),\label{lagbcm1}
\eeq
where again $\nabla_1 =\partial_1 -[A_1,]$.

\paragraph{Primary and Secondary Constraints}

The  primary constraints $\Pi_0$, $\Pi_1-\Lambda$, $\Pi_\Lambda$ are the same as in the PCM. However, the secondary constraint is now:
\beq
{\cal C} = A_0^{(2)} + \nabla_1 \Lambda \approx 0.\label{secc}
\eeq
We separate explicitly the constraint (\ref{secc}) into:
\beqz
{\cal C}^{0} =  (\nabla_1 \Lambda )^{(0)} \approx 0 \et
{\cal C}^{2} = A_0^{(2)} + (\nabla_1 \Lambda)^{(2)} \approx 0.
\eeqz
Note that the constraint $\Pi_0^{(0)}$ is  first-class since it commutes with all the constraints. The Hamiltonian density is
\beq
h = -{1\over 2} (A_0^{(2)}A_0^{(2)}-A_1^{(2)}A_1^{(2)}) + \Lambda\nabla_1 A_0  + \alpha \Pi_0 + \beta(\Pi_1 - \Lambda) + \gamma \Pi_\Lambda  + \mu {\cal C}.
\eeq
Among the Lagrange multipliers, $\mu^{(2)}$, $\alpha^{(2)}$, $\beta$, $\gamma$ are fixed and in particular $\mu^{(2)} \approx 0$. However, $\mu^{(0)}$ and $\alpha^{(0)}$ are left unfixed.

\paragraph{Hamiltonian Equations of Motion}
The Hamilton equations for $A_0$ and $A_1$ are: \label{disclageq}
\begin{align}
{dA_1 \over d\tau} &= \beta \approx \nabla_1 (A_0-\mu^{(0)}),\label{hameqbcm2} \\
{dA_0 \over d\tau} &= \alpha \approx \partial_1 A_1^{(2)}    + [ A_0^{(0)}-\mu^{(0)},A_0^{(2)}] - [ A_1^{(0)}, A_1^{(2)}] + \alpha^{(0)}. \label{hameqbcm1}
\end{align}
The reason why we concentrate on these equations of motion is the following. At the Lagrangian level, the equations of motion (\ref{lagbcm00}) and the Maurer-Cartan equation (\ref{zerogh}) are reproduced as  the zero-curvature equation for the Lax connection. The Hamiltonian equations (\ref{hameqbcm2}) and (\ref{hameqbcm1}) coincide respectively with the equations   (\ref{zerogh}) and (\ref{lagbcm00}) only when $\mu^{(0)} =0$. Therefore, one possibility is to modify the Lax connection accordingly by replacing everywhere $A_0^{(0)}$ by $A_0^{(0)}-\mu^{(0)}$. This would actually only affect the time component of the Lax connection, and is therefore irrelevant for the computation of the exchange algebra. Another possibility, explained in detail below, is to show that the condition $\mu^{(0)}=0$ simply corresponds to a gauge choice.

\paragraph{Elimination of Variables}
As for the PCM, after eliminating $\Lambda$ and $\Pi_\Lambda$, we are left with the canonical variables $(A_0,A_1,\Pi_0,\Pi_1)$. The next step is to impose strongly the set of second-class constraints
\beq
\Pi_0^{(2)}=0 \qquad \mbox{and}\qquad {\cal C}^{2} = A_0^{(2)} + ( \nabla_1 \Pi_1)^{(2)}=0. \label{edik}
\eeq
This procedure leads us to the canonical variables $(A_0^{(0)},A_1,\Pi_0^{(0)},\Pi_1)$, the constraints
\beq
\Pi_0^{(0)} \approx 0 \qquad \mbox{and} \qquad {\cal C}^0 = (\nabla_1 \Pi_1)^{(0)}\approx 0 \label{cfort}
\eeq
and the Hamiltonian density
\beq
h={1\over 2} \Bigl(   (\nabla_1\Pi_1)^{(2)} (\nabla_1\Pi_1)^{(2)} +A_1^{(2)}A_1^{(2)}\Bigr) +\alpha^{(0)} \Pi_0^{(0)} - (A_0^{(0)} - \mu^{(0)}) {\cal C}^0. \label{hlundi8}
\eeq
As already said, the constraint $\Pi_0^{(0)}$ is first-class. Before putting  the constraints (\ref{edik})  strongly to zero, the constraint ${\cal C}^0$ had only non vanishing Poisson brackets with the constraint ${\cal C}^{2}$. Therefore, it becomes also a first-class constraint in the sense of Dirac bracket, which just means that it  commutes with itself and with $\Pi_0^{(0)}$.

\paragraph{Extended Action and Gauge Invariance}

In this section, we come back to the problem of equations of motion and  study gauge invariance. To do so, we follow the approach explained in the book \cite{Henneaux} and define the so-called extended action:
\beq
S= \int d\sigma d\tau \Bigl(\Pi_1 \dot{A}_1 + \Pi_0^{(0)} \dot{A}_0^{(0)} - h\Bigr) \label{extendedaction}
\eeq
where $h$ is given by eq.(\ref{hlundi8}). The equations of motion of the Lagrange multipliers $\alpha^{(0)}$ and $\mu^{(0)}$ give the constraints (\ref{cfort}). Furthermore, these first-class constraints are associated with gauge transformations generated (in the sense of Poisson/Dirac bracket) by
 $\int d\sigma\tr(\phi \Pi_0^{(0)} + \psi {\cal C}^{0})$,
  where $\phi(\sigma,\tau),\psi(\sigma,\tau) \in {\cal G}^{(0)}$.  The corresponding  gauge transformations of the fields are:
 \beq
 \delta A_0^{(0)} = -\phi, \qquad \delta A_1 = \nabla_1 \psi, \qquad
 \delta \Pi_0^{(0)} =0, \qquad \delta \Pi_1 = [\psi,\Pi_1].\label{gg2}
 \eeq
The transformations of the  Lagrange multipliers are determined in order the action (\ref{extendedaction}) to be invariant.  We have found:
  \beq
  \delta \alpha^{(0)} = - \partial_0{\phi} \qquad \mbox{and} \qquad \delta\mu^{(0)} = -\phi -\partial_0 \psi - [\mu^{(0)} - A_0^{(0)}, \psi]. \label{gg3}
  \eeq

As already mentioned above, the action (\ref{extendedaction}) gives the equations of motion (\ref{zerogh}) and (\ref{lagbcm00}) only when $\mu^{(0)} =0$. This can be interpreted as discarding in the action (\ref{extendedaction}) the term associated with the secondary constraint ${\cal C}^0$. The reader is referred to \cite{Henneaux} for a detailed explanation. This property is however intuitively quite clear as the Legendre transform of the Lagrangian (\ref{lagbcm1}) only gives the primary constraints. In our case, this procedure can be viewed as imposing the gauge $\mu^{(0)}=0$. The residual gauge transformations  (\ref{gg2})-(\ref{gg3}) preserving that condition are such that $\phi = -\partial_0 \psi +[ A_0^{(0)}, \psi]$. The transformations (\ref{gg2}) give then
\beqz
\delta A_0^{(0)} = \partial_0 \psi -[A_0^{(0)},\psi] \et
\delta A_1 = \partial_1 \psi -[A_1,\psi].
\eeqz
Remembering that $A_0^{(2)}$ is now identified with  $ - (\nabla_1 \Pi_1)^{(2)}$, due to the second constraint in eq.(\ref{edik}), we  also find $\delta A_0^{(2)} = -[A_0^{(2)},\psi]$.  Thus, we recover the Lagrangian gauge transformations of the currents $A_\alpha$ of the coset $G/H$ model.

\subsection{Pure Spinor Formulation of $AdS_5 \times S^5$}\label{sec23}

We start the analysis of \as String theory within the pure spinor formulation\footnote{This theory contains ghosts. Here we just take the action and make the corresponding canonical analysis.}. The reason for this choice is that the pure spinor case is easier to consider than the Green-Schwarz one. Indeed, in that formulation, $\kappa$-symmetry is not present but there is an invariance under a global BRST symmetry. Therefore, for what concerns the Hamiltonian formulation, we will have to treat less constraints and the situation is  similar to the one of the previous section. We refer the reader to the appendix for some definitions and properties of the superalgebra $PSU(2,2|4)$.

\paragraph{Lagrangian Equations}
The pure spinor formulation of \as String theory is described by the Lagrangian\footnote{We use the conventions of \cite{Adam:2007ws}.} \cite{Berkovits:2000fe}:
\begin{multline*}
L= {1 \over 2} A^{(2)} \Ab^{(2)} + {1 \over 4} A^{(1)} \Ab^{(3)} + {3 \over 4} A^{(3)} \Ab^{(1)} + w \pb \lambda + \bar{w} \partial \bar{\lambda} - N \Ab^{(0)} - \bar{N} A^{(0)} - N \bar{N}.
\end{multline*}
It is written in conformal gauge\footnote{For the question related to reparametrization invariance, see \cite{Aisaka:2005vn}.}. Here, $A = -g^{-1} \partial g$ with $\partial = \partial_0 + \partial_1$ while $\Ab = - g^{-1} \pb g$ with $\pb = \partial_0 - \partial_1$. The fields $\lambda$ and $\bar{\lambda}$ are bosonic ghosts taking values in ${\cal G}^{(1)}$ and ${\cal G}^{(3)}$ respectively.  They satisfy the pure spinor conditions:
\beq
 [\lambda,\lambda]_+ =0 \qquad \mbox{and}\qquad [\bar{\lambda},\bar{\lambda}]_+=0. \label{pscond}
 \eeq
$w$ and $\bar{w}$, which will be related below to the conjugate momenta respectively of $\lambda$ and $\bar{\lambda}$, take values respectively in  ${\cal G}^{(3)}$ and ${\cal G}^{(1)}$. Finally, $N$ and $\bar{N}$ are the pure spinor currents defined by:
\beqz
N = - [w,\lambda]_+ = -w \lambda -\lambda w \et \bar{N} = - [\bar{w},\bar{\lambda}]_+= -\bar{w} \bar{\lambda} - \bar{\lambda} \bar{w}.
\eeqz
They take values in ${\cal G}^{(0)}$. The equations satisfied by the dynamical fields of this theory are the Maurer-Cartan equation,
\beq
\partial_0 A_1 = \nabla_1 A_0, \label{mcpss}
\eeq
where $\nabla_1 = \partial_1 - [A_1,]$, and the equations of motion:
\begin{align}
\t\bar{\lambda} &= [N,\bar{\lambda}],\nn\\
\t\bar{N} &= [N,\bar{N}],\nn\\
\t\Ab^{(1)} &= [N,\Ab^{(1)}]+[\bar{N},A^{(1)}],\nn\\
 \t\Ab^{(2)} &= [A^{(1)},\Ab^{(1)}]+[N,\Ab^{(2)}]+[\bar{N},A^{(2)}],\nn\\
  \t\Ab^{(3)} &= [A^{(1)},\Ab^{(2)}]+[A^{(2)},\Ab^{(1)}]+[N,\Ab^{(3)}]+[\bar{N},A^{(3)}],\nn\\
 \tb{\lambda} &= [\bar{N},\lambda],\label{eqmvtps}\\
\tb{N} &= [\bar{N},N],\nn\\
 \tb A^{(1)}&=-[A^{(2)},\Ab^{(3)}]-[A^{(3)},\Ab^{(2)}]+[N,\Ab^{(1)}]+[\bar{N},A^{(1)}],\nn\\
 \tb A^{(2)}&= -[A^{(3)},\Ab^{(3)}]+[N,\Ab^{(2)}]+[\bar{N},A^{(2)}],\nn\\
  \tb A^{(3)} &= [N,\Ab^{(3)}]+[\bar{N},A^{(3)}].\nn
 \end{align}
Here $\t = \partial -[A^{(0)},]$ and $\tb = \pb -[\Ab^{(0)},]$.

\paragraph{Primary and Secondary Constraints}

For the first-order formulation, we start accordingly with the Lagrangian\footnote{Strictly speaking, one should also introduce the terms $\phi[\lambda,\lambda]_+ + \xi \Pi_\phi$, where $\phi$ and $\xi$ are Lagrange multipliers respectively for the constraints $[\lambda,\lambda]_+ \approx 0$ and $\Pi_\phi\approx 0$. However, these two constraints are first-class and thus we choose $\phi=0$ and $\xi=0$ for simplicity. For a relevant discussion, see \cite{Bianchi:2006im}.}
\begin{align}
L =& {1 \over 2} \bigl(A_0^{(2)}A_0^{(2)} - A_1^{(2)}A_1^{(2)} \bigr) + \bigl(A_0^{(1)}A_0^{(3)} - A_1^{(1)}A_1^{(3)} \bigr) \nn\\
&+ {1\over 2} \bigl(A_0^{(1)}A_1^{(3)} - A_1^{(1)}A_0^{(3)} \bigr) + \Lambda(\partial_0 A_1 -\nabla_1 A_0) \label{lpss}\\
&+ w\partial_0 \lambda-  w\partial_1 \lambda +\bar{w} \partial_0 \bar{\lambda} + \bar{w} \partial_1 \bar{\lambda} - ( N + \bar{N}) A_0^{(0)} + (N - \bar{N}) A_1^{(0)} - N \bar{N}\nn
\end{align}
where the dynamical fields are now $(A_0,A_1,\Lambda,\lambda,\bar{\lambda})$. The Hamiltonian analysis goes as follows. As for the two previous cases,  the   set of primary constraints is:
\beqz
\Pi_0 \approx 0, \qquad \Pi_1 - \Lambda \approx 0, \qquad \Pi_\Lambda \approx 0.
\eeqz
Concerning the pure spinor fields, $-w_\alpha$ and $-\bar{w}_\beta$ are respectively the conjugate momenta of $\lambda^\alpha$ and $\bar{\lambda}^\beta$. The P.B. of the canonical variables are given in section \ref{appa2} of the appendix. The Hamiltonian density $h= \Pi_1 \partial_0 A_1 - w  \partial_0 \lambda- \bar{w}  \partial_0 \bar{\lambda} -L$ is then:
\begin{align}
h = &-{ 1 \over 2} \bigl(A_0^{(2)}A_0^{(2)} - A_1^{(2)}A_1^{(2)} \bigr) - \bigl(A_0^{(1)}A_0^{(3)} - A_1^{(1)}A_1^{(3)} \bigr)
- {1\over 2} \bigl(A_0^{(1)}A_1^{(3)} - A_1^{(1)}A_0^{(3)} \bigr)  \nn \\
&+ w\partial_1 \lambda  - \bar{w} \partial_1 \bar{\lambda} + ( N + \bar{N}) A_0^{(0)} - (N - \bar{N}) A_1^{(0)} + N \bar{N} + \Lambda\nabla_1 A_0 \label{hfirstps} \\
&+\alpha \Pi_0 + \beta(\Pi_1 - \Lambda) + \gamma \Pi_\Lambda +\mu {\cal C} \nn
\end{align}
where we have already included the secondary constraint
\beq
{\cal C} = A_0^{(1)} + A_0^{(2)} + A_0^{(3)} - \half(A_1^{(1)} - A_1^{(3)}) + \nabla_1 \Lambda - N - \bar{N}. \label{cps}
\eeq
We then find:
\begin{align}
\{ \Pi_0, H \} &= {\cal C} - \mu^{(1)} - \mu^{(2)} -\mu^{(3)},\nn\\
 \{  \Pi_1 - \Lambda, H \} &= - \gamma -[\Lambda, A_0 - \mu] -(A_1^{(1)} + A_1^{(2)} + A_1^{(3)} ) \nn\\
 &\qquad \quad+ \half(A_0^{(1)} - A_1^{(3)}) + N - \bar{N},
\label{forgamma}\\
 \{  \Pi_\Lambda , H\} &= -\nabla_1(A_0 -\mu) + \beta,\nn\\
\{ {\cal C}, H \} &= \alpha^{(1)} + \alpha^{(2)} + \alpha^{(3)} - \Psi\nn
\end{align}
where
\begin{multline}
\Psi = \half(\beta^{(1)} - \beta^{(3)}) +[\beta, \Lambda] - \nabla_1 \gamma\\ + \partial_1(N - \bar{N}) -[N, A_0^{(0)} - \mu^{(0)}-A_1^{(0)}] -[\bar{N}, A_0^{(0)} - \mu^{(0)} + A_1^{(0)}]. \label{defpsips}
\end{multline}
Therefore, preservation of the constraints under time evolution implies in particular:
\begin{align}
\mu^{(1)} &\approx 0, \qquad \mu^{(2)} \approx 0, \qquad \mu^{(3)} \approx 0,\label{aaaa}\\
\beta &\approx \nabla_1(A_0-\mu^{(0)}), \label{aaab}\\
\Psi &\approx \alpha^{(1)} + \alpha^{(2)} +\alpha^{(3)}.\label{232}
\end{align}
Using the definition (\ref{defpsips}) of $\Psi$, the result (\ref{forgamma}) for $\gamma$ and the results (\ref{aaaa})-(\ref{aaab}), one finds that $\Psi^{(0)} \approx 0$. This means that there is no further constraint. The equations (\ref{defpsips}) and (\ref{232}) enable then to determine  $\alpha^{(1)}$, $\alpha^{(2)}$ and $\alpha^{(3)}$. Therefore, all the Lagrange multipliers are fixed except $\alpha^{(0)}$ and $\mu^{(0)}$. As expected, this is the same situation as in the previous section.

\paragraph{Hamiltonian Equations of Motion}

The Hamiltonian equations of motion are:
\begin{align*}
{ dA_1 \over d\tau } &= \beta \approx \nabla_1(A_0-\mu^{(0)}),\\
{ d A_0 \over d\tau} &= \alpha,\\
{ d N \over d\tau} &\approx \partial_1 N + [A_0^{(0)} - \mu^{(0)} - A_1^{(0)} + \bar{N}, N],\\
{d\bar{N} \over d\tau} &\approx -\partial_1 \bar{N} + [A_0^{(0)} - \mu^{(0)} + A_1^{(0)} + N, \bar{N}].
\end{align*}
To be more explicit, one needs to determine  $\alpha^{(1)}$, $\alpha^{(2)}$ and $\alpha^{(3)}$. This is a lengthy but straightforward computation. In a similar way to the situation examined previously for the bosonic $G/H$ Coset model, one finds that the Hamiltonian equations of motion coincide with the Lagrangian ones, provided that $A_0^{(0)}$ is replaced everywhere  by $A_0^{(0)} -\mu^{(0)}$.

\paragraph{Elimination of Variables}

As usual now, we first eliminate $\Lambda$ and $\Pi_\Lambda$. We are left with the canonical variables $(A_0,A_1,\Pi_0,\Pi_1,\lambda,w,\bar{\lambda},\bar{w})$ together with the constraints $\Pi_0 \approx 0$ and
\begin{align}
{\cal C}^0 &= (\nabla_1 \Pi_1)^{(0)} - (N + \bar{N})  \approx 0,\nn\\
 {\cal C}^1 &= A_0^{(1)} - {1 \over 2} A_1^{(1)} +(\nabla_1 \Pi_1)^{(1)} \approx 0, \label{idps1}\\
{\cal C}^2 &=  A_0^{(2)} + (\nabla_1 \Pi_1)^{(2)} \approx 0, \label{idps2}\\
 {\cal C}^3 &= A_0^{(3)} + {1 \over 2} A_1^{(3)} +  (\nabla_1 \Pi_1)^{(3)}\approx 0.\label{idps3}
\end{align}
We eliminate then the variables $A_0^{(1,2,3)}$ and $\Pi_0^{(1,2,3)}$ by putting strongly to zero the system $(\Pi_0^{(1,2,3)},{\cal C}^{1,2,3})$ of second-class constraints.
\paragraph{Summary}

The first-order Hamiltonian formulation of pure spinor \as String theory consists of the canonical variables $(A_0^{(0)},\Pi_0^{(0)},A_1, \Pi_1, \lambda, w, \bar{\lambda}, \bar{w})$, whose fundamental Poisson brackets are given in the appendix, and the  first-class constraints
\beq
\Pi_0^{(0)} \approx 0 \et {\cal C}^0 = (\nabla_1 \Pi_1)^{(0)} - (N + \bar{N})  \approx 0. \label{idps0}
\eeq
In particular, $\{ {\cal C}^0_\uu(\sigma), {\cal C}^0_\dd(\sigma') \} = [C_{\uu\dd}^{(00)}, {\cal C}^0_\dd]\dss \approx 0$.  Finally, starting from eq.(\ref{hfirstps}), one finds the Hamiltonian density:
\begin{align*}
h &= \half \Bigl[ (\nabla_1 \Pi_1)^{(2)} (\nabla_1 \Pi_1)^{(2)} + A_1^{(2)} A_1^{(2)} \Bigr] \nn \\
 &+ (\nabla_1 \Pi_1)^{(1)} (\nabla_1 \Pi_1)^{(3)} + \half \Bigl[  (\nabla_1 \Pi_1)^{(1)}  A_1^{(3)} - (\nabla_1 \Pi_1)^{(3)}  A_1^{(1)} \Bigr]
+{3 \over 4}  A_1^{(1)}  A_1^{(3)} \nn\\
&+ w\partial_1 \lambda -\bar{w}\partial_1 \bar{\lambda} - (N - \bar{N}) A_1^{(0)} + N \bar{N}+ \alpha^{(0)} \Pi_0^{(0)} - (A_0^{(0)} - \mu^{(0)}) {\cal C}^0.
\end{align*}

\subsection{Green-Schwarz Formulation of  $AdS_5 \times S^5$} \label{sec24}

We refer the reader to the appendix for some definitions and results related to the superalgebra $PSU(2,2|4)$. Our starting point is the Lagrangian \cite{Metsaev:1998it,Roiban:2000yy}
\beqz
L = -{1 \over 2}  \left[ \gamma^{\alpha\beta} (g^{-1}\partial_\alpha g)^{(2)} (g^{-1}\partial_\beta g)^{(2)} + \kappa \epsilon^{\alpha\beta} (g^{-1}\partial_\alpha g)^{(1)} (g^{-1}\partial_\beta g)^{(3)}\right].
\eeqz
Here, the group element $g(\sigma,\tau)$ belongs to $\mbox{PSU}(2,2|4)$; we use the convention $\epsilon^{01}=\epsilon^{\tau\sigma}=1$; $\gamma^{\alpha\beta}$ is the Weyl-invariant combination of the world-sheet metric with $\mbox{det} \gamma =-1$; taking the supertrace is not explicitly written.  Finally, we have taken a general coefficient in front of the Wess-Zumino term. Remember however that invariance under $\kappa$-symmetry imposes $\kappa =\pm 1$.

\paragraph{Lagrangian Equations}

The equations satisfied by the current $A_\alpha = - g^{-1} \partial_\alpha g$ are the Maurer-Cartan equation
\beq
\partial_0 A_1 = \nabla_1 A_0, \label{mcmardi}
\eeq
where we have defined the covariant derivative $\nabla_1 = \partial_1 - [A_1,]$, and the equation of motion
\beq
\partial_\alpha S^\alpha - [A_\alpha, S^\alpha]=0 \label{eqmvt}
\eeq
with $S^\alpha =   \gamma^{\alpha\beta} A_\beta^{(2)} - {1 \over 2} \kappa \epsilon^{\alpha\beta} (A_\beta^{(1)} - A_\beta^{(3)})$. The equation  (\ref{eqmvt}) does not give anything on ${\cal G}^{(0)}$ and gives respectively for ${\cal G}^{(2)}$, ${\cal G}^{(1)}$ and ${\cal G}^{(3)}$:
\begin{align}
\partial_\alpha\bigl( \gamma^{\alpha \beta} A_\beta^{(2)} \bigr) - \gamma^{\alpha\beta} [A_\alpha^{(0)}, A_\beta^{(2)} ] + {1 \over 2} \kappa \epsilon^{\alpha\beta}\Bigl( [A_\alpha^{(1)}, A_\beta^{(1)} ] - [A_\alpha^{(3)}, A_\beta^{(3)} ] \Bigr) &=0,\label{eqgr2}\\
 [P_-^{\beta\alpha}A_\alpha^{(3)}, A_\beta^{(2)} ] &=0, \label{eqgr1}\\
[P_+^{\beta\alpha}A_\alpha^{(1)}, A_\beta^{(2)} ]  &=0, \label{eqgr3}
\end{align}
where the Maurer-Cartan equation (\ref{mcmardi}) has been used. We have  also introduced
\beqz
P_\pm^{\alpha\beta} \equiv {1 \over 2} (\gamma^{\alpha\beta} \pm \kappa \epsilon^{\alpha\beta}).
\eeqz
These operators are orthogonal projectors when $\kappa = \pm 1$. Let us make here an important remark. The equation of motion (\ref{eqgr2}) is of the form $\partial_0 A_0^{(2)} + ... =0$. However, when they are written in terms of the currents, the equations of motion on the odd gradings do not contain any derivative. This will have some consequence below.

To these equations, one has to add the Virasoro constraints
\beq
T_{\alpha\beta} = \str(A_\alpha^{(2)} A_\beta^{(2)}) - \half \gamma_{\alpha\beta} \gamma^{\rho\sigma} \str(A_\rho^{(2)} A_\sigma^{(2)}) \approx 0. \label{virc2}
\eeq
The strategy we will follow concerning these constraints is the following.
 First of all, we will not introduce conjugate momenta for the metric because the Hamiltonian would then become rather cumbersome. Thus, the Virasoro constraints will be imposed "by hand". We will also consider at the beginning the theory without imposing the Virasoro constraints. It will only be imposed later in the process, when some of the redundant variables will already have been  eliminated. This procedure is correct because the matrix of the P.B. of the constraints we will strongly put to zero remains invertible, even when Virasoro constraints are taken into account.

\paragraph{Primary and Secondary Constraints}

Let us start therefore with the Lagrangian
\beq
 L = - {1 \over 2}  \left[ \gamma^{\alpha\beta} A_\alpha^{(2)} A_\beta^{(2)} + \kappa \epsilon^{\alpha\beta} A_\alpha^{(1)} A_\beta^{(3)}\right] + \Lambda(\partial_0 A_1 -\nabla_1 A_0) \label{lgs}
\eeq
for the dynamical variables $(A_0,A_1,\Lambda)$ and do the Legendre transform. As usual, the set of primary constraints is:
\beqz
\Pi_0\approx 0,\qquad \Pi_1 - \Lambda \approx 0, \qquad \Pi_\Lambda \approx 0.
\eeqz
The Hamiltonian density is then
\begin{multline}
h ={1  \over 2}  \left[ \gamma^{\alpha\beta} A_\alpha^{(2)} A_\beta^{(2)} + \kappa \epsilon^{\alpha\beta} A_\alpha^{(1)} A_\beta^{(3)}\right] + \Lambda(\nabla_1 A_0) \\+ \alpha \Pi_0 + \beta(\Pi_1 - \Lambda) + \gamma \Pi_\Lambda + \mu {\cal C} \label{hhgs}
\end{multline}
where the secondary constraint
\beq
{\cal C} = - \gamma^{0\alpha} A_\alpha^{(2)} + {\kappa \over 2} \bigl(A_1^{(1)} - A_1^{(3)}\bigr) + \nabla_1 \Lambda \approx 0 \label{avant243}
\eeq
follows from imposing the stability  of the primary constraint $\Pi_0$. We then have $\{\Pi_0,H\} \approx \gamma^{00} \mu^{(2)}$ such that $\mu^{(2)} \approx 0$. Stability of the constraint $\Pi_1 - \Lambda$ leads to the result
\beqz
\gamma = - \gamma^{1\alpha} A_\alpha^{(2)} -[\Lambda, A_0 -\mu] - {\kappa \over 2} \bigl( A_0^{(1)} - \mu^{(1)} \bigr) +{\kappa \over 2} \bigl(A_0^{(3)} - \mu^{(3)}\bigr) + \gamma^{01} \mu^{(2)}.
\eeqz
For $\Pi_\Lambda$, one finds as usual
\beq
\beta = \nabla_1 (A_0 - \mu).  \label{asusual}
 \eeq
 For the stability of the constraint ${\cal C}$, we have to take into account the explicit time dependence in the metric. We find:
\begin{align*}
{d {\cal C}\over d\tau} &= {\partial_0 {\cal C}} + \{ {\cal C}, H \}
= - (\partial_0 \gamma^{0\alpha}) A_\alpha^{(2)}- \gamma^{00} \alpha^{(2)} - \Psi
\end{align*}
where we have defined:
\beqz
\Psi =  \gamma^{01} \beta^{(2)} + {\kappa \over 2} (\beta^{(3)} -\beta^{(1)}) + [\beta, \Lambda] - \nabla_1 \gamma.
\eeqz
The condition $(d{\cal C}/d\tau)\approx 0$ requires therefore that
\beq
\Psi  \approx - (\partial_0 \gamma^{0\alpha}) A_\alpha^{(2)} - \gamma^{00} \alpha^{(2)}. \label{dcdtau}
\eeq
After some algebra, we find $\Psi^{(0)} \approx 0$, which is fine, and:
\begin{align}
\Psi^{(1)} &\approx [A_0^{(2)}, -\gamma^{00} \mu^{(3)} + 2 P_-^{0\alpha} A_\alpha^{(3)}] +[A_1^{(2)}, -(\gamma^{01} + \kappa)\mu^{(3)} + 2 P_-^{1\alpha} A_\alpha^{(3)}], \label{psigs1}\\
\Psi^{(3)}&\approx [A_0^{(2)} , -\gamma^{00} \mu^{(1)} + 2 P_+^{0\alpha} A_\alpha^{(1)}] +[A_1^{(2)}, -(\gamma^{01}-\kappa)\mu^{(1)} + 2 P_+^{0\alpha} A_\alpha^{(1)}]\nn
\end{align}
For $\Psi^{(2)}$ we have
\begin{align}
\Psi^{(2)} &\approx \partial_1(\gamma^{1\alpha} A_\alpha^{(2)}) + \gamma^{01} \partial_1 A_0^{(2)} \nn\\
&- 2 \gamma^{01} [ A_1^{(0)}, A_0^{(2)}] - \gamma^{11} [A_1^{(0)}, A_1^{(2)}] + \gamma^{00} [A_0^{(2)}, A_0^{(0)} - \mu^{(0)}] \label{refps2}\\
& + (\kappa -\gamma^{01}) [A_1^{(3)}, A_0^{(3)} -\mu^{(3)}] - (\kappa + \gamma^{01}) [A_1^{(1)}, A_0^{(1)} - \mu^{(1)}].\nn
\end{align}
Let us examine first $\Psi^{(1)}$ and the corresponding condition on $\mu^{(3)}$. We must have $\Psi^{(1)} \approx 0$ (see eq.(\ref{dcdtau})). Using the general result
\beq
{\gamma^{01}+\kappa \over \gamma^{00}} P_-^{0\alpha}X_\alpha  = P_-^{1\alpha}X_\alpha + { 1- \kappa^2 \over 2 \gamma^{00}} X_1, \label{prokappa}
\eeq
we rewrite the relation (\ref{psigs1}) as:
\beqz
\Psi^{(1)} \approx 2[P_+^{0\alpha}A_\alpha^{(2)}, {2 \over \gamma^{00}}P_-^{0\alpha} A_\alpha^{(3)} - \mu^{(3)}] + {\kappa^2-1\over \gamma^{00}} [A_1^{(2)},  A_1^{(3)}].
\eeqz
Thus, when $\kappa^2 =1$, which corresponds precisely to the condition in order to have $\kappa$-symmetry, there is no further constraint. We fix from now on $\kappa=1$. We then have:
\beq
\mu^{(3)}= {2 \over \gamma^{00}} P_-^{0\alpha} A_\alpha^{(3)} + \widetilde{\mu}^{(3)} \qquad \mbox{with} \qquad [ P_+^{0\alpha} A_\alpha^{(2)}, \widetilde{\mu}^{(3)}] \approx 0. \label{mumu3}
\eeq
We obtain similarly:
\beq
\mu^{(1)} ={2 \over \gamma^{00}} P_+^{0\alpha} A_\alpha^{(1)} + \widetilde{\mu}^{(1)} \qquad \mbox{with} \qquad
[P_-^{0\alpha} A_\alpha^{(2)}, \widetilde{\mu}^{(1)}] \approx 0. \label{mumu1}
\eeq
At this point, one should not forget that the Virasoro constraints have also to be taken into account. This is why we have included the terms $\widetilde{\mu}^{(1,3)}$. This freedom is indeed present when the Virasoro constraints (\ref{virc2}) are also imposed and is related to $\kappa$-symmetry. For our purpose, which is the computation of the exchange algebra, it is not necessary to exactly determine this freedom. It is however clear that the analysis goes along the lines of the one presented for instance in \cite{Arutyunov} for the $AdS_4 \times CP^3$ case (see eq.(3.6) and (4.5) of that reference).

To summarize, the Lagrange multipliers  $\alpha^{(0)}$, $\alpha^{(1)}$, $\alpha^{(3)}$,  $\mu^{(0)}$ are unfixed and  $\mu^{(1)}$ and $\mu^{(3)}$  are only partially fixed.

\paragraph{Partial Gauge-Fixing}

As the constraints $\Pi_0^{(1)}  \approx 0$ and $\Pi_0^{(3)} \approx 0$ are first-class, they generate gauge transformations. We introduce  the gauge-fixing conditions
\beq
{\cal D}^1 = P_+^{0\alpha} A_\alpha^{(1)} \approx 0 \et
{\cal D}^3 = P_-^{0\alpha} A_\alpha^{(3)}\approx 0. \label{dd13}
\eeq
In conformal gauge, such conditions have been considered in \cite{Grigoriev:2007bu} to partially fix $\kappa$-symmetry. In the present case, they are natural to introduce if we take into account the expressions of $\mu^{(1)}$ and $\mu^{(3)}$ and the general discussion page \pageref{disclageq} on Hamiltonian equations of motion. Furthermore, it is immediate to see that they are suitable  gauge-fixing conditions as they form a set of  second-class constraints  with $\Pi_0^{(3)}$ and $\Pi_0^{(1)}$. For instance,   $\{ {\cal D}^1_\uu(\sigma), \Pi_{0\dd}^{(3)}(\sigma') \} = (1/2)C_{\uu\dd}^{(13)} \gamma^{00} \dss$. For the time evolution of ${\cal D}^1$ and ${\cal D}^3 $, we have:
\begin{align*}
{d {\cal D}^1 \over d\tau} &\approx \half( \partial_0 \gamma^{00})  A_0^{(1)} + \half( \partial_0 \gamma^{01})  A_1^{(1)} +\half \gamma^{00} \alpha^{(1)} +\half (\gamma^{01} +1)\beta^{(1)},\\
{d {\cal D}^3 \over d\tau} &\approx \half( \partial_0 \gamma^{00})  A_0 ^{(3)}  - \half (\partial_0 \gamma^{01})  A_1 ^{(3)} +\half \gamma^{00} \alpha^{(3)} + \half (\gamma^{01}-1) \beta^{(3)}.
\end{align*}
Imposing $(d  {\cal D}^{1,3} / d\tau) \approx 0$ and using eq.(\ref{asusual}) gives $\alpha^{(1)}$ and $\alpha^{(3)}$ in terms of the Lagrange multipliers $\mu^{(0)}$, $\widetilde{\mu}^{(1)}$ and  $\widetilde{\mu}^{(3)}$. Thus, at this level, the only freedom left is in the Lagrange multipliers $\alpha^{(0)}$, $\mu^{(0)}$, $\widetilde{\mu}^{(1)}$ and  $\widetilde{\mu}^{(3)}$.

\paragraph{Hamiltonian Equations of Motion}

Let us look at the Hamiltonian equations of motion. As in the previous cases, we first find
\beq
{ dA_1 \over d\tau} =\beta \approx\nabla_1 (A_0-\mu) \label{mcgs}
\eeq
for $A_1$.
For $A_0$, a first result is:
\beq
{ dA_0 \over d\tau} = \alpha \approx \alpha^{(0)} + \alpha^{(1)} + \alpha^{(3)} -{1 \over \gamma^{00}} \Psi^{(2)} - {1 \over \gamma^{00}} (\partial_0 \gamma^{0\alpha}) A_\alpha^{(2)}, \label{seceqgs}
\eeq
where we have used the relation between $\alpha^{(2)}$ and $\Psi^{(2)}$ (see eq.(\ref{dcdtau})).  We consider this equation grading by grading. On ${\cal G}^{(0)}$, we obtain
\beq
{dA_0^{(0)} \over d\tau}= \alpha^{(0)}. \label{hgr0}
\eeq
Then, using the expression (\ref{refps2}) of $\Psi^{(2)}$, the first equation of motion (\ref{mcgs}) and the constraints (\ref{dd13}), the projection on ${\cal G}^{(2)}$ of eq.(\ref{seceqgs}) can be rewritten as:
\begin{align}
\partial_\alpha(\gamma^{\alpha\beta} A_\beta^{(2)}) &= \gamma^{10}[ A_1^{(0)}, A_0^{(2)}] + \gamma^{01} [A_0^{(0)} - \mu^{(0)}, A_1^{(2)}] \nn \\
& + \gamma^{00}[A_0^{(0)} - \mu^{(0)}, A_0^{(2)}] + \gamma^{11} [ A_1^{(0)}, A_1^{(2)}] \label{hgr2}\\
&+ [A_1^{(1)}, A_0^{(1)} - \widetilde{ {\mu}}^{(1)}] - [A_1^{(3)},A_0^{(3)} -\widetilde{{\mu}}^{(3)}].\nn
\end{align}
For the odd gradings, computing the equations of motion for $A_0^{(1)}$  and $A_0^{(3)}$, one recovers the conditions $(d{\cal D}^1/d\tau) =0$ and $(d{\cal D}^3/d\tau) =0$.

Let us compare these results with the Lagrangian equations (\ref{mcmardi}) and (\ref{eqgr2})-(\ref{eqgr3}). First of all, one recovers the Maurer-Cartan equation (\ref{mcmardi})  only when $\mu^{(0)} =0$, $\widetilde{\mu}^{(1)} = 0$ and $\widetilde{\mu}^{(3)} = 0$. The same property holds for the comparison between eq.(\ref{hgr2}) and eq.(\ref{eqgr2}). The equations (\ref{eqgr1}) and (\ref{eqgr3}) corresponding to the odd gradings are recovered, but as a consequence of the partial gauge-fixing conditions (\ref{dd13}). This is so because $P_\pm^{0\alpha} X_\alpha =0$ implies $P_\pm^{1\alpha} X_\alpha =0$ (see eq.(\ref{prokappa})).

\paragraph{Elimination of Variables}

As usual, we first eliminate $\Lambda$ and $\Pi_\Lambda$. Then, we put strongly to zero the set of second-class constraints $\Pi_0^{(2)} =0$ and   ${\cal C}^{(2)} =0$. We interpret this process as elimination of the variables $\Pi_0^{(2)}$ and $A_0^{(2)}$. In particular, putting strongly ${\cal C}^{(2)} =0$ means that  (see eq.(\ref{avant243})):
\beq
A_0^{(2)} = {1 \over \gamma^{00}} \bigl[ (\nabla_1 \Pi_1)^{(2)} - \gamma^{01} A_1^{(2)} \bigr]. \label{c2pgs}
\eeq
We eliminate then $A_0^{(1,3)}$ (by using eq.(\ref{dd13})) and $\Pi_0^{(1,3)}$. This procedure leads to the canonical variables $(A_0^{(0)},A_1,\Pi_0^{(0)},\Pi_1)$ and the constraints
\begin{align}
\Pi_0^{(0)} \approx 0, \qquad {\cal C}^0 = (\nabla_1 \Pi_1)^{(0)} \approx 0,\nn\\
{\cal C}^{1} = (\nabla_1 \Pi_1)^{(1)}  + {1  \over 2}  A_1^{(1)} \approx 0, \qquad  {\cal C}^{3} =(\nabla_1 \Pi_1)^{(3)} -  {1  \over 2}  A_1^{(3)} \approx 0, \label{c1c3tobeput}\\
T_{\alpha\beta} = \str(A_\alpha^{(2)} A_\beta^{(2)}) - \half \gamma_{\alpha\beta} \gamma^{\rho\sigma} \str(A_\rho^{(2)} A_\sigma^{(2)}) \approx 0,\nn
\end{align}
where we have now included the Virasoro constraints. The  constraints $\Pi_0^{(0)}$ and ${\cal C}^0$  are first-class. In the second line, as usual in the Green-Schwarz formulation, there is a mixing of first-class and second-class constraints, due to the Virasoro constraints. We have\footnote{As these P.B. are  ultralocal,  $\delta_{\sigma\sigma'}$ is not indicated.}
\begin{align*}
\{ \C^0_\uu, \C^i_\dd \} &=  \bigl[C_{\uu\dd}^{(00)}, \C^i_\dd\bigr] \approx 0, \quad i=0,1,3  \et \{ \C^1_\uu, \C^3_\dd \} = \bigl[C_{\uu\dd}^{(13)}, \C^0_\dd\bigr] \approx 0,\\
\{ \C^1_\uu, \C^1_\dd \} &=    \bigl[C_{\uu\dd}^{(13)}, (\nabla_1 \Pi_1)_\dd^{(2)} + A_{1 \dd}^{(2)}\bigr] ,\\
\{ \C^3_\uu, \C^3_\dd \} & =   \bigl[C_{\uu\dd}^{(31)}, (\nabla_1 \Pi_1)_\dd^{(2)} - A_{1 \dd}^{(2)}\bigr].
\end{align*}
Finally, the Hamiltonian density is:
\begin{multline}
\widetilde{h} = -{1 \over 2 \gamma^{00}} (\nabla_1 \Pi_1)^{(2)} (\nabla_1 \Pi_1)^{(2)}  - { 1 \over 2 \gamma^{00}} A_1^{(2)} A_1^{(2)} - {1 \over \gamma^{00}} A_1^{(1)} A_1^{(3)} + {\gamma^{01} - 1 \over \gamma^{00}}  (\nabla_1 \Pi_1)^{(1)} A_1^{(3)} \\+ {\gamma^{01} +1 \over \gamma^{00}} (\nabla_1 \Pi_1)^{(3)}A_1^{(1)}
-(A_0^{(0)} - \mu^{(0)}) {\cal C}^0 + \alpha^{(0)} \Pi_0^{(0)} + \widetilde{\mu}^{(1)} {\cal C}^3 +
\widetilde{\mu}^{(3)} {\cal C}^1. \label{htildegs}
\end{multline}

\paragraph{Comment} Let us comment here another approach, which is sometimes used to derive the phase space structure of coset models. As explained page \pageref{motupage}, this approach is based on the relation $A_0 = \nabla_1^{-1}(\partial_0 A_1)$. One problem of this  approach is that it does not give any information at all on certain variables like $A_0^{(0,1,3)}$. This is actually not a problem for the computation done in \cite{Das:2004hy} as the spatial "Lagrangian" Lax component does not depend on those variables. In fact, we will also not need any information on those variables for the computation of the P.B. of the "Hamiltonian" spatial Lax component. This is however a problem if one wants to go further in the Hamiltonian analysis and consider for instance gauge-fixing conditions like (\ref{dd13}).

\section{Exchange Algebra}\label{sec3}

\subsection{Hamiltonian Lax Connection} \label{sec32}

\paragraph{Pure Spinor Formulation}

The Lagrangian Lax connection for the pure spinor formulation has been determined in \cite{Vallilo:2003nx}. Here we use a similar parameterization as the one in  \cite{Mikhailov:2007mr} and introduce:
\begin{align*}
 {\cal L}(z)  &= \bigl(A^{(0)} +N -z^4 N\bigr) + zA^{(1)} + z^2 A^{(2)} + z^3 A^{(3)},\\
 \bar{{\cal L}}(z)  &= \bigl(\Ab^{(0)} +\bar{N} -z^{-4} \bar{N}\bigr) +  z^{-3} \Ab^{(1)} +  z^{-2} \Ab^{(2)} + z^{-1}\Ab^{(3)},
 \end{align*}
 with the same notations as in section \ref{sec23} and with $z$ the spectral parameter. The zero-curvature equation
 \beqz
 \bar{\partial} {\cal L} - \partial\bar{ {\cal L}} -[\bar{{\cal L}}, {\cal L} ]=0
 \eeqz
implies the Maurer-Cartan equation (\ref{mcpss}) and the equations of motion\footnote{Strictly speaking, for what concerns the ghosts, it only implies the equations of motion for $N$ and $\bar{N}$.} (\ref{eqmvtps}). It can be easily checked that
\beq
\Omega\bigl({\cal L}(z)\bigr) = {\cal L}(iz) \label{condgradingl}
\eeq
and similarly for $\bar{\cal L}(z)$, where $\Omega$ is the Lie algebra homomorphism related to the grading (see eq.(\ref{grading}) in appendix). We are interested in the spatial component i.e. in
\begin{multline}
 \half\bigl[ {\cal L}(z) - \bar{\cal L}(z)\bigr]
= \half \Bigl[ 2A_1^{(0)} + (1-z^4)N -(1-z^{-4}) \bar{N} +A_1^{(1)}(z+z^{-3}) + A_0^{(1)}(z-z^{-3}) \\
\quad + A_1^{(2)}(z^2+z^{-2}) + A_0^{(2)}(z^2-z^{-2})+A_1^{(3)}(z^3+z^{-1}) + A_0^{(3)}(z^3-z^{-1})\Bigr]. \label{v815}
\end{multline}
The corresponding expression at the Hamiltonian level is found by using the constraints (\ref{idps1})-(\ref{idps3}), that we have put strongly to zero. As already mentioned in the introduction, we  add to this Lax component (\ref{v815}) a term proportional to the constraint ${\cal C}^0$ defined by eq.(\ref{idps0}). In principle, the coefficient multiplying this constraint is completely arbitrary. It just needs to satisfy the condition (\ref{condgradingl}) related to the grading. However, to simplify the discussion, we  fix it to a particular value and indicate in \S\ref{sec33} what happens for other values of this coefficient. Let us therefore add the term  $\rho(z){\cal C}^0$ to the
component (\ref{v815}) with $\rho(z) = (1/2)(1-z^4)$. The corresponding result is called the Hamiltonian spatial Lax component and denoted by $\LL_1(z)$:
\begin{multline}
\LL_1(z)  =  A_1^{(0)} + a A_1^{(1)} + b A_1^{(2)} + c A_1^{(3)} \\
 + \rho (\nabla_1\Pi_1)^{(0)}+\gamma (\nabla_1\Pi_1)^{(1)} +\beta  (\nabla_1\Pi_1)^{(2)}+ \alpha  (\nabla_1\Pi_1)^{(3)}
 + \bar{\xi} \bar{N}
 \label{finalmbis}
 \end{multline}
 with
 \begin{align*}
 a(z) &= {1\over 4}(3z + z^{-3}),\qquad &\qquad b(z) &=\half(z^2+z^{-2}),\\
 c(z) &= {1\over 4}(z^3 + 3z^{-1}),\qquad &\qquad  \gamma(z) &= \half(z^{-3}-z),\\
 \beta(z) &= \half(z^{-2}-z^2), \qquad &\qquad \alpha(z) &= \half(z^{-1}-z^3),\\
 \rho(z) &= \half(1-z^4), \qquad &\qquad \bar{\xi}(z) &= \half(z^{-4}+z^4-2).
 \end{align*}

\paragraph{Green-Schwarz Formulation}\label{sec31}

The Lagrangian Lax connection has been determined in \cite{Bena:2003wd}. Here, we follow a similar parametrization as the one in \cite{Beisert:2005bm}. The spatial component ${\cal \widetilde{L}}_1(z)$ of this connection is:
\beqz
 {\cal \widetilde{L}}_1(z) = A_1^{(0)} + z A_1^{(1)} + z^{-1} A_1^{(3)} + {1 \over 2} (z^2 + z^{-2}) A_1^{(2)} + {1 \over 2}(-z^2 + z^{-2})\gamma^{0\alpha}A_\alpha^{(2)} .
 \eeqz
 where $z$ is the spectral parameter. At the Hamiltonian level, this corresponds to
 \beqz
 {\cal \widetilde{L}}_1(z) = A_1^{(0)} + z A_1^{(1)} + z^{-1} A_1^{(3)} + {1 \over 2} (z^2 + z^{-2}) A_1^{(2)} + {1 \over 2}(-z^2 + z^{-2})(\nabla_1 \Pi_1)^{(2)}
 \eeqz
 where we have simply used the relation (\ref{c2pgs}).  We first add to this Lax component terms proportional to the constraints (\ref{c1c3tobeput}),
 \beq
 {\cal C}^{1} = (\nabla_1 \Pi_1)^{(1)}  + {1  \over 2}  A_1^{(1)} \approx 0 \et  {\cal C}^{3} =(\nabla_1 \Pi_1)^{(3)} -  {1  \over 2}  A_1^{(3)} \approx 0. \label{tralala}
 \eeq
Again, the coefficients multiplying these constraints are in principle completely arbitrary. They just need to satisfy the condition (\ref{condgradingl}) related to the grading. We will however fix them to a particular value. This value is chosen such that the coefficients multiplying $ (\nabla_1 \Pi_1)^{(1)} $ and $ (\nabla_1 \Pi_1)^{(3)}$ in the new Lax component are the same as in the pure spinor formulation (\ref{finalmbis}). Similarly, we also add the term $(1/2)(1-z^4){\cal C}^0$ where ${\cal C}^0 = (\nabla_1 \Pi_1)^{(0)}$ in the Green-Schwarz formulation. Therefore, we define
 \beqz
 {\cal L}_1(z) = {\cal \widetilde{L}}_1(z)  + \half (z^{-3}-z) {\cal C}^1 + \half(z^{-1}-z^3) {\cal C}^3 + \half(1-z^4){\cal C}^0.
 \eeqz
Using the relations (\ref{tralala}), we actually find that ${\cal {L}}_1(z)$ is exactly the same as the Hamiltonian Lax connection (\ref{finalmbis})  of the pure spinor formulation, up to the term proportional to the ghost current ${\bar{N}}$. Note that the world-sheet metric has not been fixed and that it does not explicitly appear in the expression of ${\cal L}_1(z)$, when this component is written in terms of the phase space variables.

\subsection{Poisson Brackets of the spatial Lax Component}\label{sec32bis}

As the Hamiltonian spatial Lax components are the same in both Green-Schwarz and pure spinor formulations up to the term proportional to the ghosts, and with the same P.B. for the canonical variables, we do the computation of the P.B. $\{ \LL_\uu(\sigma,z_1), \LL_\dd(\sigma',z_2)\}$
including the ghost term. The result for the Green-Schwarz case is then simply recovered by taking $\bar{\xi}=0$. We note  $\LL(z) \equiv \LL_1(z)$. All terms appearing in this Poisson bracket  are straightforward to compute and   will be listed below. The real problem encountered is that all ultra-local terms can be rewritten in two different ways due to the identities (\ref{c12c21}). Therefore, we need a strategy to organize this computation. It consists simply in starting from the desired form of the result, namely the $r/s$ form, and in determining if it corresponds to what we actually obtain. Let us define therefore $\LL_\pm = \LL_\uu(z_1) \pm \LL_\dd(z_2)$ and recall that the $r/s$ form is:
\begin{multline}
\{ \LL_\uu(\sigma,z_1), \LL_\dd(\sigma',z_2)\} = [r_{\uu\dd}(z_1,z_2), \LL_+]\dss -[s_{\uu\dd}(z_1,z_2),\LL_-] \dss \\- 2 s_{\uu\dd}(z_1,z_2) \pdss \label{ourconv2}
\end{multline}
where $r$ is antisymmetric and $s$ symmetric.  The non-ultra-local terms are easily identified. This leads to the result:
 \begin{multline}
2s_{\uu\dd}(z_1,z_2) = \Bigl[(\rho_1 + \rho_2)C_{\uu\dd}^{(00)}+ \bigl(b_1 \beta_2 + b_2 \beta_1\bigr) C_{\uu\dd}^{(22)} \\+ \bigl(a_1 \alpha_2 + \gamma_1 c_2\bigr) C_{\uu\dd}^{(13)}
+ \bigl(a_2 \alpha_1 + \gamma_2 c_1\bigr) C_{\uu\dd}^{(31)}\Bigr]. \label{sfound}
 \end{multline}
 Let us then search $r_{\uu\dd}$ as:
 \beq
 r_{\uu\dd}= A_{12} C_{\uu\dd}^{(00)} + B_{12} C_{\uu\dd}^{(22)} + D_{12} C_{\uu\dd}^{(13)} -D_{21} C_{\uu\dd}^{(31)} \label{redik}
 \eeq
 with $A$ and $B$ antisymmetric, i.e. $A_{12} = A(z_1,z_2) =-A(z_2,z_1)$ and similarly for $B$. For completeness, we have  $D_{12} = D(z_1,z_2)$  and $D_{21} = D(z_2,z_1)$. Define in a similar way
 \beq
 s_{\uu\dd}=  \widetilde{A}_{12} C_{\uu\dd}^{(00)}+ \widetilde{B}_{12} C_{\uu\dd}^{(22)} + \widetilde{D}_{12} C_{\uu\dd}^{(13)} +\widetilde{D}_{21} C_{\uu\dd}^{(31)}, \label{sput}
 \eeq
 according to the result (\ref{sfound}). We work out the sought  $r/s$ form. We use the fact that all the ultra-local terms can be cast in the form $[C_{\uu\dd}^{(i4-i)},X_\dd]$. Therefore,  we first write
 \beq
 [r_{\uu\dd}, \LL_+] -[s_{\uu\dd}, \LL_-] = [r_{\uu\dd}-s_{\uu\dd}, \LL_\uu] +[r_{\uu\dd}+s_{\uu\dd}, \LL_\dd]. \label{rpsp}
 \eeq
We keep then the last term in the r.h.s. of (\ref{rpsp}) as it has the right structure. To treat the $r-s$ terms, the following properties are used:
 \begin{align*}
 [C_{\uu\dd}^{(00)}, \LL_\uu(z_1)]=&-[C_{\uu\dd}^{(00)},\LL_\dd^{(0)}(z_1)]
 -[C_{\uu\dd}^{(13)},\LL_\dd^{(1)}(z_1)]\\
 &-[C_{\uu\dd}^{(22)},\LL_\dd^{(2)}(z_1)]
 -[C_{\uu\dd}^{(31)},\LL_\dd^{(3)}(z_1)],\\
  [C_{\uu\dd}^{(13)}, \LL_\uu(z_1)]=&-[C_{\uu\dd}^{(13)},\LL_\dd^{(0)}(z_1)]
 -[C_{\uu\dd}^{(22)},\LL_\dd^{(1)}(z_1)]\\
& -[C_{\uu\dd}^{(31)},\LL_\dd^{(2)}(z_1)]
 -[C_{\uu\dd}^{(00)},\LL_\dd^{(3)}(z_1)],\\
 [C_{\uu\dd}^{(22)}, \LL_\uu(z_1)]=&-[C_{\uu\dd}^{(22)},\LL_\dd^{(0)}(z_1)]
 -[C_{\uu\dd}^{(31)},\LL_\dd^{(1)}(z_1)]\\
&-[C_{\uu\dd}^{(00)},\LL_\dd^{(2)}(z_1)]
 -[C_{\uu\dd}^{(13)},\LL_\dd^{(3)}(z_1)],\\
 [C_{\uu\dd}^{(31)}, \LL_\uu(z_1)]=&-[C_{\uu\dd}^{(31)},\LL_\dd^{(0)}(z_1)]
 -[C_{\uu\dd}^{(00)},\LL_\dd^{(1)}(z_1)]\\
& -[C_{\uu\dd}^{(13)},\LL_\dd^{(2)}(z_1)]
 -[C_{\uu\dd}^{(22)},\LL_\dd^{(3)}(z_1)].
 \end{align*}
All these identities are obtained as a consequence of eq.(\ref{c12c21}). Let us insist that in the r.h.s. of these identities, the spectral parameter is $z_1$.  For each different projection of the quadratic Casimir, we collect now the diverse terms appearing in (\ref{rpsp}). This gives the list of the terms we shall have in the P.B. $\{\LL_\uu,\LL_\dd\}$ if it is of the $r/s$ form. It is now time to compare this list with the terms we do have in the expression of this P.B. For each projection of the Casimir we look grading by grading and term by term. This is summarized for $C_{\uu\dd}^{(00)}$ in the following table:
$$
\begin{array}{|l|l|c|}
\hline
C_{\uu\dd}^{(00)}&\mbox{Expected}&\mbox{Found}\\
\hline
&&\\
A_{1\dd}^{(0)}&2\widetilde{A}_{12}&\rho_1+\rho_2\\
A_{1\dd}^{(1)}&(A_{12}+\widetilde{A}_{12})a_2 +(D_{21}+\widetilde{D}_{21})a_1&\gamma_2+a_2 \rho_1\\
A_{1\dd}^{(2)}&(A_{12}+\widetilde{A}_{12})b_2 -(B_{12} -\widetilde{B}_{12})b_1&\beta_2+b_2 \rho_1\\
A_{1\dd}^{(3)}&(A_{12}+\widetilde{A}_{12})c_2 -(D_{12} - \widetilde{D}_{12})c_1&\alpha_2+c_2 \rho_1\\
(\nabla_1 \Pi_1)_\dd^{(0)}&(A_{12}+\widetilde{A}_{12})\rho_2 -(A_{12}-\widetilde{A}_{12})\rho_1&\rho_1\rho_2\\
(\nabla_1 \Pi_1)_\dd^{(1)}&(A_{12}+\widetilde{A}_{12})\gamma_2 + (D_{21} + \widetilde{D}_{21})\gamma_1 &\rho_1\gamma_2\\
(\nabla_1 \Pi_1)_\dd^{(2)}&(A_{12}+\widetilde{A}_{12})\beta_2 -(B_{12}-\widetilde{B}_{12})\beta_1&\rho_1 \beta_2\\
(\nabla_1 \Pi_1)_\dd^{(3)}&(A_{12}+\widetilde{A}_{12})\alpha_2 -(D_{12}-\widetilde{D}_{12})\alpha_1 &\rho_1 \alpha_2\\
\bar{N}_\dd&(A_{12}+\widetilde{A}_{12})\bar{\xi}_2 - (A_{12}-\widetilde{A}_{12})\bar{\xi}_1&-\bar{\xi}_1\bar{\xi}_2\\
\hline
\end{array}
$$
The fifth row of this table together with the result (\ref{sfound}) simply give:
\beqz
A_{12} = \half {\rho_1^2 + \rho_2^2 \over \rho_1-\rho_2}.
\eeqz
Then, the sixth to eighth rows enable us to compute respectively $B$ and $D$. At this stage, $r$ is therefore completely determined. We will give its expression shortly. It remains however to check that the conditions associated with the other rows are also satisfied. This is indeed the case. We go on with all the other projections but  it is now just a matter of checking the tables that we have put in section \ref{tablesap} of the appendix. We find perfect agreement for all these conditions. We can therefore summarize what we have found.

\paragraph{Summary} The P.B. of the spatial component of the Hamiltonian Lax connection (\ref{finalmbis}) for both Green-Schwarz and pure spinor formulations has the following form:
\begin{multline}
\{ \LL_\uu(\sigma,z_1), \LL_\dd(\sigma',z_2)\} = [r_{\uu\dd}(z_1,z_2), \LL_+]\dss -[s_{\uu\dd}(z_1,z_2),\LL_-] \dss \\- 2 s_{\uu\dd}(z_1,z_2) \pdss
\label{theresult}
\end{multline}
with:
\begin{multline}
s_{\uu\dd}(z_1,z_2) = {1\over 4}\bigl(2-z_1^4-z_2^4\bigr) C_{\uu\dd}^{(00)}
+ {1 \over 4}\bigl( z_1^{-2}z_2^{-2} - z_1^2 z_2^2\bigr) C_{\uu\dd}^{(22)}\\
+{1 \over 4}\bigl(  z_1^{-3} z_2^{-1} - z_1 z_2^3\bigr) C_{\uu\dd}^{(13)}  + {1 \over 4}\bigl(z_2^{-3} z_1^{-1} - z_2 z_1^3\bigr) C_{\uu\dd}^{(31)}\label{vensmat}
\end{multline}
and
\begin{multline}
r_{\uu\dd}(z_1,z_2) = -{1 \over 4(z_1^4-z_2^4)} \Biggl[\Bigl((1-z_1^4)^2+(1-z_2^4)^2 \Bigr) C_{\uu\dd}^{(00)} \\
+\Bigl((z_1^2-z_1^{-2})^2 +(z_2^2-z_2^{-2})^2\Bigr)\Bigl( z_1z_2^3 C_{\uu\dd}^{(13)} + z_1^3z_2 C_{\uu\dd}^{(31)} + z_1^2z_2^2 C_{\uu\dd}^{(22)}\Bigr) \Biggr]. \label{rmatdim003}
\end{multline}

\subsection{Discussion}\label{sec34}\label{sec33}

\paragraph{Jacobi Identity and Yang-Baxter Equations}

As indicated page \pageref{YB2}, a sufficient condition for the Jacobi identity to be satisfied is that $r$ and $s$ are solutions of the extended Yang-Baxter equation (\ref{YB2}). To prove that this property holds in the present case, we follow the analysis of \cite{Mail2} and write $r$ and $s$ as\footnote{I thank J.~M.~Maillet for pointing out that method.}
\beq
r_{\uu\dd} + s_{\uu\dd} =  f(z_1) {\Pi_{\uu\dd}(z_1,z_2) \over z_1^4 -z_2^4}\et
r_{\uu\dd} - s_{\uu\dd} =  f(z_2) {\widetilde{\Pi}_{\uu\dd}(z_2,z_1) \over z_1^4 -z_2^4} \label{bepulse}
\eeq
with
\begin{align}
\Pi_{\uu\dd}(z_1,z_2) &= C_{\uu\dd}^{(00)} + z_1^{-2}z_2^2 C_{\uu\dd}^{(22)} + z_1^{-3}z_2^3 C_{\uu\dd}^{(13)} + z_1^{-1}z_2 C_{\uu\dd}^{(31)},\\
\widetilde{\Pi}_{\uu\dd}(z_2,z_1) &=  C_{\uu\dd}^{(00)} + z_2^{-2}z_1^2 C_{\uu\dd}^{(22)}  + z_2^{-1}z_1 C_{\uu\dd}^{(13)} + z_2^{-3}z_1^3 C_{\uu\dd}^{(31)},\\
f(z)&=-{1 \over 2} (1-z^4)^2. \label{bepulsebis}
\end{align}
Note that $\widetilde{\Pi}_{\uu\dd}(z_2,z_1) = P \Pi_{\uu\dd}(z_2,z_1)P$ where  $P(A\otimes B) P=  (-)^{|A||B|} B \otimes A$ for any matrices $A$ and $B$. Defining then
\beqz
X_{\uu\dd\underline{\mathbf{3}}} = [r_{\uu{\underline{{\mathbf{3}}}}} + s_{\uu{\underline{{\mathbf{3}}}}},
r_{\uu\dd}- s_{\uu\dd}] + [r_{\dd{\underline{{\mathbf{3}}}}} + s_{\dd{\underline{{\mathbf{3}}}}}, r_{\uu\dd}+ s_{\uu\dd}] +[r_{\dd{\underline{{\mathbf{3}}}}}+s_{\dd{\underline{{\mathbf{3}}}}}, r_{\uu{\underline{{\mathbf{3}}}}}+s_{\uu{\underline{{\mathbf{3}}}}}],
\eeqz
one finds that
\beqz
X_{\uu\dd\ttr} = {f(z_1) f(z_2) \over (z_1^4-z_2^4)(z_1^4-z_3^4)(z_2^4-z_3^4)} Y_{\uu\dd\ttr}
\eeqz
with
\begin{multline*}
Y_{\uu\dd\ttr} = (z_2^4-z_3^4)[ \Pi_{\uu\ttr}(z_1,z_3), \widetilde{\Pi}_{\uu\dd}(z_2,z_1)]  +
(z_1^4-z_3^4)[ \Pi_{\dd\ttr}(z_2,z_3), \Pi_{\uu\dd}(z_1,z_2)]\\
+ (z_1^4-z_2^4)[ \Pi_{\dd\ttr}(z_2,z_3), \Pi_{\uu\ttr}(z_1,z_3)].
\end{multline*}
The first contribution to  $Y_{\uu\dd\ttr}$ is a sum of terms proportional to $[C^{(i\,4-i)}_{\uu\dd}, C^{(j \,4-j)}_{\uu\ttr}]$. Using the relations\footnote{These relations are obtained from eq.(\ref{eq1}). We recall that the tensor product is graded (see eq.(\ref{tgra1})).}
\begin{align*}
[C_{\dd\ttr}^{(k\,4-k)}, C_{\uu\dd}^{(i \,4-i)}] &=- [ C_{\uu\ttr}^{(k\,4-k)}, C_{\uu\dd}^{(4-k+i \;k-i)}],\\
[C_{\dd\ttr}^{(k\,4-k)},C_{\uu\ttr}^{(i\,4-i)}] &=-[C_{\uu\dd}^{(4-k\,k)},C_{\uu\ttr}^{(k+i\;4-k-i)}],
\end{align*}
it is possible to also write the two other contributions to  $Y_{\uu\dd\ttr}$ as  a sum of terms proportional to $[C^{(i\,4-i)}_{\uu\dd}, C^{(j \,4-j)}_{\uu\ttr}]$. It is then straightforward to collect all these terms and to show that their sum vanishes.

Note that it is possible to show, by using the same method as above, that the matrix $r$ is not a solution of the Yang-Baxter equation (\ref{yb}).

\paragraph{Exchange Algebra}

As already explained in the introduction, a consequence of the result (\ref{theresult}) is that the
conserved charges of this theory are in involution. Furthermore, the monodromy matrix defined by eq.(\ref{defmon}) satisfies the classical exchange algebra:
\beqz
\{ T_\uu(z_1), T_\dd(z_2) \} = [r_{\uu\dd}, T_1(z_1)T_2(z_2)] + T_\uu(z_1) s_{\uu\dd} T_\dd(z_2) - T_\dd(z_2) s_{\uu\dd} T_\uu(z_1).
\eeqz
It is understood here that the regularization introduced in \cite{Maillet:1985ek} is used.

\paragraph{Effect of first-class Constraints and Comparison with \cite{Mikhailov:2007eg}}

Let us start from the Hamiltonian Lax component (\ref{finalmbis}) and discuss the effect of varying the coefficient $\rho$ multiplying the first-class constraint ${\cal C}^0$.

Consider first the case $\rho=0$. For the pure spinor case, this corresponds to work with the Lagrangian Lax component, which has been used by A. Mikhailov and S. Sch\"afer-Nameki in  \cite{Mikhailov:2007eg}. Making the same analysis as above, we have found that the P.B.  has not exactly the $r/s$ form. Indeed, there is an  additional term, proportional to ${\cal C}^0$:
\beq
\Bigl[\gamma_1\alpha_2 C_{\uu\dd}^{(13)} + \gamma_2\alpha_1  C_{\uu\dd}^{(31)} +\beta_1\beta_2 C_{\uu\dd}^{(22)},   {\cal C}^0_\dd\Bigr]\dss. \label{additional1}
\eeq
Furthermore, the  matrices $r^0$ and $s^0$ corresponding to this choice $\rho=0$  differ from the ones in  (\ref{rmatdim003}) and (\ref{vensmat}), but only by terms proportional to $C_{\uu\dd}^{(00)}$. More precisely, using the definitions given by eq.(\ref{redik}) and (\ref{sput}), we have:
\beqz
A_{\uu\dd}^0 = - {(1-z_1^4)(1-z_2^4) \over 2(z_1^4-z_2^4)} C_{\uu\dd}^{(00)} \et \widetilde{A}^0_{\uu\dd} = 0.
\eeqz

Apart from an inessential global factor, $s^0$ is the same matrix as the one in \cite{Mikhailov:2007eg} while $r^0$ is the opposite of the one found in \cite{Mikhailov:2007eg}. However, the origin of this discrepancy is probably only a matter of convention for the definition of the $r/s$ form. Indeed, the eq.(2.35) in \cite{Mikhailov:2007eg}, which is used for the extended Yang-Baxter equation, corresponds to a convention where the sign of $r$ (or equivalently of $s$) is flipped with respect to our convention (\ref{ourconv2}) and the corresponding extended Yang-Baxter equation (\ref{YB2}).
 
The additional term (\ref{additional1}) is absent in \cite{Mikhailov:2007eg}. However, this comes {\em a priori} from the fact that the observables considered in \cite{Mikhailov:2007eg} are gauge invariant. Indeed, this statement would be in agreement with the property that ${\cal C}^0$ generates gauge transformations. An explicit comparison would therefore be possible by computing the P.B. of gauge invariant observables. This additional term (\ref{additional1})  has  to be taken into account for the Jacobi identity. The actual presence of this term explains why the matrices $r^0$ and $s^0$ do not satisfy the extended Yang-Baxter equation (\ref{YB2}) but a generalization of this equation (see \cite{Mikhailov:2007eg} for details). 

Another Hamiltonian Lax connection leads to a $r/s$ form for its P.B. It corresponds to the choice  $\rho(z) = (1/2)(z^{-4}-1)$. Then, in the expression (\ref{finalmbis}), the term $\bar{\xi}\bar{N}$ has to be replaced by $\xi N$ with $\xi(z) = -(1/2)(z^{-4}+z^4-2)$. The corresponding matrices $r$ and $s$ are obtained from the ones in  (\ref{rmatdim003}) and (\ref{vensmat}) by changing the terms proportional to $C_{\uu\dd}^{(00)}$:
\begin{align*}
s_{\uu\dd}:&\quad&{1 \over 4}(2-z_1^{4}-z_2^{4})C_{\uu\dd}^{(00)} &\quad&\rightarrow&\quad& -{1 \over 4}(2-z_1^{-4}-z_2^{-4})C_{\uu\dd}^{(00)},\\
r_{\uu\dd}:&\quad&-{ (1-z_1^{4})^2 +(1-z_2^{4})^2 \over 4 (z_1^{4}-z_2^{4})}C_{\uu\dd}^{(00)} &\quad&\rightarrow&\quad&{ (1-z_1^{-4})^2 +(1-z_2^{-4})^2 \over 4 (z_1^{-4}-z_2^{-4})}C_{\uu\dd}^{(00)}.
\end{align*}

This discussion  also illustrates  the general comment made in the introduction, page \pageref{pagecomment}. Indeed, we concretely see on these two examples that the $r/s$ form is preserved, at least on the constraint surface, when the coefficient proportional to the first-class constraint ${\cal C}^0$ has been changed.

\paragraph{Effect of second-class Constraints}

What does happen now for the Green-Schwarz case if we do not include the terms proportional to the  constraints ${\cal C}^0$, ${\cal C}^1$ and ${\cal C}^3$ in the Hamiltonian Lax connection, i.e. if we work with the Lagrangian Lax connection ? This corresponds in (\ref{finalmbis}) to:
\beqz
\rho=0, \qquad \gamma=0, \qquad \alpha =0, \qquad a(z)=z,\qquad c(z)=z^{-1},
\eeqz
the other coefficients, $b$ and $\beta$, being unchanged. Then, the analysis goes as follows. First of all, we also obtain the additional term (\ref{additional1}). For the matrix $s_{\uu\dd}$,  $\widetilde{A}$ and $\widetilde{D}$ vanish (see eq.(\ref{sfound})). For $r_{\uu\dd}$, working out the terms proportional to $A_1^{(2)}$ and $(\nabla_1\Pi_1)^{(2)}$, one obtains that $D_{12}=0$ and
$$
A_{12} = {\beta_1\beta_2 \over \beta_1 b_2 -\beta_2 b_1}, \qquad B_{12} = \widetilde{B}_{12} + {1\over b_1}(A_{12}b_2 -\beta_2).
$$
Looking then at the terms proportional to $A_1^{(1)}$ and $(\nabla_1\Pi_1)^{(1)}$, one finds a difference between the expected and found terms. Furthermore, this difference is not proportional to the constraint ${\cal C}^1$. Therefore, the systematic method used here shows that the P.B. of the Bena-Polchinski-Roiban spatial Lax component is not of the $r/s$ form, even when it is evaluated on the constraint surface. This is in agreement with the result obtained in \cite{Das:2004hy}. As part of the constraints ${\cal C}^1$ and ${\cal C}^3$ are second-class, this shows that changing the coefficients multiplying second-class constraints affects the form of the Poisson brackets.

\paragraph{Link between Green-Schwarz and pure Spinor Formulations}

We have found the same classical exchange algebra for both Green-Schwarz (G.S.) and pure spinor  (P.S.)  descriptions of \as String theory. One would like however to see explicitly and at the level of this first-order Hamiltonian formulation, that these two descriptions are equivalent. For that, one has to  completely gauge fix $\kappa$-symmetry in the G.S. formulation, in the spirit of what has been done in \cite{Berkovits:2004tw}. However, as we will discuss it in the conclusion, gauge fixing is a difficult task within the first-order Hamiltonian formulation. Therefore, we only make a much simpler observation. Consider the G.S. action in conformal gauge and the P.S. action. A simple inspection of these two actions (see eq.(\ref{lpss}) and (\ref{lgs})) shows that these formulations will "meet" if one does simultaneously the following\footnote{It is also possible to see it at the level of the equations of motion but it requires more work as one has to use the Maurer-Cartan equation.} \cite{Vallilo:2003nx,Mikhailov:2007mr}:

\noindent - For the G.S. formulation: Impose the conditions:
\beq
A_0^{(1)} = A_1^{(1)} \et A_0^{(3)} = - A_1^{(3)}. \label{meet}
\eeq
- For the P.S. formulation: Discard the ghosts and impose the same conditions (\ref{meet}).

Concerning the G.S. formulation, one has already imposed the conditions (\ref{meet}) in section \ref{sec24}: they correspond indeed to the conditions (\ref{dd13}) in the special case of conformal gauge\footnote{With $\gamma^{00}=-1$.}.  For the P.S. formulation, remember that the meaning of the variables $A_0^{(1)}$ and $A_0^{(3)}$ in Hamiltonian formulation is given by the equations (\ref{idps1}) and (\ref{idps3}), corresponding to  constraints we have strongly put to zero. Therefore, the conditions (\ref{meet}) should be rather read as
\beq
\half A_1^{(1)} + (\nabla_1 \Pi_1)^{(1)} =0 \et -\half A_1^{(3)} + (\nabla_1 \Pi_1)^{(3)} =0. \label{ouf}
\eeq
But these are precisely the constraints (\ref{c1c3tobeput}) encountered in the Green-Schwarz formulation.  It would remain to compute for the P.S. case the new Hamiltonian preserving the constraints (\ref{ouf}). However, as the equations of motion are the same, it is clear that one shall recover the Hamiltonian density (\ref{htildegs}).

\section{Conclusion}

We conclude by first making some comments in the framework of the more general problem of non-ultra-local terms.

{\em A priori},  the first-order Hamiltonian formulation used in this article only holds in the classical case so far. However, the next step would be to directly find the quantum analogue of the classical exchange algebra. For instance, when this algebra has the form $\{ T_\uu, T_\dd \} = [r_{\uu\dd}, T_\uu T_\dd]$, with $r$ satisfying the classical Yang-Baxter equation, this is a sign that in the quantum case one shall have
\beq
R_{12}T_1 T_2 =T_2T_1R_{12} \label{qyb}
\eeq
where $R$ satisfies the quantum Yang-Baxter equation \cite{yang-baxter}. However, in the present case, and as already mentioned, the P.B. of the monodromy matrix are not well defined. Even if there exists a regularization\footnote{See also \cite{Duncan:1989vg} for another approach.} of these P.B. \cite{Maillet:1985ek}, the Jacobi identity is only "weakly" satisfied, which is clearly a problem for finding the quantum analogue of the classical exchange algebra. A generalization of the quadratic algebra (\ref{qyb}) has been proposed, on general grounds, in \cite{Freidel:1991jx}. It is simply $A_{12}T_1 B_{12} T_2 = T_2 C_{12} T_1 D_{12}$. However, $A$ and $D$ satisfy the quantum Yang-Baxter equation in the framework of \cite{Freidel:1991jx}. This means that their classical analogue satisfy the classical Yang-Baxter equation. But the matrix $r$ we have found does not satisfy the classical Yang-Baxter equation.  It is therefore not clear at the moment what is the quantum version of (\ref{theresult}) and the only available results so far consist in the approach developed in \cite{Mikhailov:2007eg} and the subsequent conjecture made there.

A question related to the present discussion is: what is the link between the  matrix $\Pi$  we have found and the classical $r$-matrices found in \cite{Torrielli:2007mc} ? Two other related questions concern the algebraic origins of $\Pi$ and of the Hamiltonian Lax component. We expect that both can be understood by generalizing to the  $PSU(2,2|4)$ case the construction presented in \cite{Mail2}.

\medskip

In the case of the principal chiral model, a way to deal with non-ultra-local terms corresponds to the Faddeev-Reshetikhin approach \cite{Faddeev:1985qu}. In the context of \as, it has been considered in \cite{Klose:2006dd}. It would be very interesting to develop this approach within the first-order Hamiltonian presented in this article.

This is however the Zamolodchikov-Zamolodchikov approach \cite{Zamolo}, i.e. the determination of the factorized $S$-matrix from its symmetries and properties, which is the most successful in the context  of \as \cite{refall}. For that reason, it would be desirable to study the uniform light-cone gauge considered in  \cite{Arutyunov:2004yx}, \cite{Frolov:2006cc}. However, a strong limitation of the first-order Hamiltonian formulation is that there is no direct access to the P.B. of the group element with the currents. This information is however needed as the gauge-fixing conditions defining the uniform light-cone-gauge are expressed  in terms of the currents and the group element. In fact, they even involve explicit use of coordinates. It is therefore not obvious at all that the advantage of    only  dealing  with the currents can be kept in the process of fully gauge-fix the theory.

The first-order Hamiltonian formulation might however be more useful for the study of the 2d duality of \as \cite{Ricci:2007eq} related to the dual superconformal symmetry of scattering amplitudes in $N=4$ super-Yang-Mills theory \cite{Drummond:2006rz}.

\medskip

Let us however conclude in an optimistic way by making the following general remark concerning non-ultra-local terms. One should perhaps turn this discussion the other way round. In the long-term, the study of \as String theory might lead to a better understanding of how to generally  deal with non-ultra-local terms. The fact that N.~Dorey and B.~Vicedo have been able to construct action-angle variables from the finite gap solutions data and for a subsector of \as \cite{Dorey:2006mx,Vicedo:2008jk} might be considered as an encouraging sign  since the Jacobi identity is fully satisfied by action-angle variables.

\paragraph{Acknowledgements}

 I thank  N.~Beisert, F.~Delduc, S. Frolov, F.~Gieres, V. Kazakov, Y.~Kazama, T.~McLoughlin, J.M.~Maillet, H.~Samtleben, M. Staudacher and B.~Vicedo for  discussions and the Albert-Einstein-Institut for its kind hospitality. This work is also partially supported by Agence Nationale de la Recherche under the contract ANR-07-CEXC-010.

\appendix

\section{Appendix} \label{app0}
\subsection{Definitions and Notations} \label{appa1}

The superalgebra $SU(2,2|4)$ admits a $\mathbb{Z}_4$  grading induced by some Lie algebra homomorphism $M\to \Omega(M)$ (see for instance \cite{Frolov:2006cc} for details). This means that it is  decomposed as a vector space into the direct sum ${\cal G}^{(0)} \oplus{\cal G}^{(1)} \oplus {\cal G}^{(2)} \oplus{\cal G}^{(3)}$. Each subspace is an eigenspace of $\Omega$ i.e., for any $M^{(k)} \in {\cal G}^{(k)}$:
\beq
\Omega(M^{(k)}) = i^k M^{(k)}.\label{grading}
\eeq
We note generically $t_A \in {\cal G}$ and for each grading $t_a \in {\cal G}^{(0)}$, $t_\alpha \in {\cal G}^{(1)}$, $t_i \in {\cal G}^{(2)}$, $t_\beta \in {\cal G}^{(3)}$. We then have
\beqz
\eta_{AB} \equiv \str(t_At_B), \quad \eta_{BA} = (-)^{|A|} \eta_{AB}, \quad \eta^{AB}\eta_{BC} = \delta^A_C
\eeqz
where $\str$ is the supertrace and $|A|=0,1$ respectively for even and odd gradings. For $M=M^A t_A$ we define $M_A = \str(T_A M)$. The graded commutator $[,]$ is defined as
\beqz
[t_A,t_B] = t_A t_B - (-)^{|A||B|} t_B t_A = f_{AB}^{\phantom{AB}C} t_C,
\eeqz
where the structure constants satisfy
\begin{eqnarray*}
f_{AB}^{\phantom{AB}D} \eta_{DC} &=& -(-)^{|A||B|} f_{BA}^{\phantom{BA}D} \eta_{DC} = -(-)^{|B||C|} f_{AC}^{\phantom{AC}D} \eta_{DB}.
\end{eqnarray*}

\paragraph{Tensor Product and Quadratic Casimir}

We use a   graded tensor product
\beq
(t_A  \otimes t_B) (t_C \otimes t_D) = (-)^{|B||C|} (t_A t_C) \otimes (t_B t_D). \label{tgra1}
\eeq
The quadratic Casimir is defined by:
\begin{align*}
C_{\uu\dd} &= \eta^{AB} t_A \otimes t_B= \eta^{ab} t_a \otimes t_b + \eta^{\alpha\beta} t_\alpha \otimes t_\beta + \eta^{ij} t_i \otimes t_j + \eta^{\beta\alpha} t_\beta \otimes t_\alpha,\\
 &= C_{\uu\dd}^{(00)} + C_{\uu\dd}^{(13)} + C_{\uu\dd}^{(22)} + C_{\uu\dd}^{(31)}.
\end{align*}
It satisfies the property
\beq
[C_{\uu\dd}, M_\uu] = -[C_{\uu\dd}, M_\dd]. \label{eq1}
\eeq
The relation (\ref{eq1}) can be projected on the different gradings:
\begin{align}
[C_{\uu\dd}^{(i \, 4-i)}, M_\dd^{(i+j)} ] &=-[ C_{\uu\dd}^{(4-j \, j)}, M_\uu^{(i+j)} ]. \label{c12c21}
\end{align}

\subsection{Poisson Brackets}\label{appa2}

Let $\Pi_1 = \Pi_1^A t_A$ be the conjugate momentum of  $A_1 = A_1^A t_A$. The canonical P.B. is
\beq
\{ A_{1\uu}(\sigma), \Pi_{1\dd}(\sigma') \} = \Ca \dss. \label{eq2}
\eeq
In components, this corresponds to $\{ A_1^A(\sigma), \Pi_1^B(\sigma') \} = \eta^{AB} \dss$.

\paragraph{Poisson Brackets for Ghosts in the Pure Spinor Formulation}

The P.B. given below are ultralocal. Therefore we do not write explicitly $\dss$.
\begin{align*}
\{ \lambda_\uu, w_\dd \} &= C^{(13)}_{\uu\dd}, &\qquad \{ \bar{\lambda}_\uu, \bar{w}_\dd \} &= C^{(31)}_{\uu\dd},\\
\{N_\uu,N_\dd\} &= -[C_{\uu\dd}^{(00)},N_\dd], &\qquad
\{\bar{N}_\uu,\bar{N}_\dd\} &= -[C_{\uu\dd}^{(00)},\bar{N}_\dd]. 
\end{align*}

\subsection{Constraints and Dirac Bracket}

For a constrained system:

A constraint is first-class if its Poisson brackets with all the other constraints vanish on the constraint surface.

A set $({\cal C}_\alpha)$ of constraints is a set of second-class constraints if the matrix $M_{\alpha\beta}$ formed by the P.B. $\{{\cal C}_\alpha,{\cal C}_\beta\}$ is invertible. The Dirac bracket associated with this set of second-class constraints  is defined by
\begin{multline}
\{f(\sigma),g(\sigma')\}_D = \{f(\sigma),g(\sigma')\} - \int d\sigma_1 d\sigma_2 \{ f(\sigma), {\cal C}_\alpha(\sigma_1)\}(M^{-1})_{\alpha\beta}(\sigma_1,\sigma_2) \times\\
\{ {\cal C}_\beta(\sigma_2),g(\sigma')\}. \label{defdiracapp}
\end{multline}
It satisfies $\{f, {\cal C}_\alpha \}_D=0$ for any function $f$ and enables therefore to put the constraints ${\cal C}_\alpha$ strongly to zero.

\subsection{Tables relevant for Section \ref{sec32bis}} \label{tablesap}

$$
\begin{array}{|l|l|c|}
\hline
C_{\uu\dd}^{(13)}&\mbox{Expected}&\mbox{Found}\\
\hline
\phantom{a}&\phantom{b}&\phantom{c}\\
A_{1\dd}^{(0)}&2\widetilde{D}_{12}&a_1\alpha_2+c_2\gamma_1\\
A_{1\dd}^{(1)}&(D_{12}+\widetilde{D}_{12})a_2 -(A_{12}-\widetilde{A}_{12})a_1&\gamma_1+a_1 \rho_2\\
A_{1\dd}^{(2)}&(D_{12}+\widetilde{D}_{12})b_2 +(D_{21}+\widetilde{D}_{21})b_1&a_1\gamma_2 +a_2\gamma_1\\
A_{1\dd}^{(3)}&(D_{12}+\widetilde{D}_{12})c_2 -(B_{12} - \widetilde{B}_{12})c_1&a_1\beta_2 + b_2 \gamma_1\\
(\nabla_1 \Pi_1)_\dd^{(0)}&(D_{12}+\widetilde{D}_{12})\rho_2 - (D_{12}-\widetilde{D}_{12})\rho_1&\gamma_1 \alpha_2\\
(\nabla_1 \Pi_1)_\dd^{(1)}&(D_{12}+\widetilde{D}_{12})\gamma_2 -(A_{12}-\widetilde{A}_{12})\gamma_1 &\gamma_1 \rho_2\\
(\nabla_1 \Pi_1)_\dd^{(2)}&(D_{12}+\widetilde{D}_{12})\beta_2 +(D_{21}+\widetilde{D}_{21})\beta_1&\gamma_1\gamma_2\\
(\nabla_1 \Pi_1)_\dd^{(3)}&(D_{12}+\widetilde{D}_{12})\alpha_2 -(B_{12}-\widetilde{B}_{12})\alpha_1 &\gamma_1\beta_2\\
\bar{N}_\dd&(D_{12}+\widetilde{D}_{12})\bar{\xi}_2 - (D_{12} - \widetilde{D}_{12})\bar{\xi}_1& 0\\
\hline
\end{array}
$$\label{tabke}
$$
\begin{array}{|l|l|c|}
\hline
C_{\uu\dd}^{(22)}&\mbox{Expected}&\mbox{Found}\\
\hline
&&\\
A_{1\dd}^{(0)}&2\widetilde{B}_{12}&b_1\beta_2+b_2\beta_1\\
A_{1\dd}^{(1)}&(B_{12}+\widetilde{B}_{12})a_2 -(D_{12}-\widetilde{D}_{12})a_1&b_1\alpha_2+c_2\beta_1\\
A_{1\dd}^{(2)}&(B_{12}+\widetilde{B}_{12})b_2 -(A_{12}-\widetilde{A}_{12})b_1&\beta_1+b_1\rho_2\\
A_{1\dd}^{(3)}&(B_{12}+\widetilde{B}_{12})c_2 +(D_{21} + \widetilde{D}_{21})c_1&a_2\beta_1 + b_1 \gamma_2\\
(\nabla_1 \Pi_1)_\dd^{(0)}&(B_{12}+\widetilde{B}_{12})\rho_2- (B_{12}-\widetilde{B}_{12})\rho_1&\beta_1\beta_2\\
(\nabla_1 \Pi_1)_\dd^{(1)}&(B_{12}+\widetilde{B}_{12})\gamma_2 -(D_{12}-\widetilde{D}_{12})\gamma_1 &\beta_1\alpha_2\\
(\nabla_1 \Pi_1)_\dd^{(2)}&(B_{12}+\widetilde{B}_{12})\beta_2 -(A_{12}-\widetilde{A}_{12})\beta_1&\beta_1\rho_2\\
(\nabla_1 \Pi_1)_\dd^{(3)}&(B_{12}+\widetilde{B}_{12})\alpha_2 +(D_{21}+\widetilde{D}_{21})\alpha_1 &\gamma_2\beta_1\\
\bar{N}_\dd&(B_{12}+\widetilde{B}_{12})\bar{\xi}_2 -(B_{12}-\widetilde{B}_{12})\bar{\xi}_1&0\\
\hline
\end{array}
$$
$$
\begin{array}{|l|l|c|}
\hline
C_{\uu\dd}^{(31)}&\mbox{Expected}&\mbox{Found}\\
\hline
&&\\
A_{1\dd}^{(0)}&2\widetilde{D}_{21}&\alpha_1a_2+c_1\gamma_2\\
A_{1\dd}^{(1)}&(-D_{21}+\widetilde{D}_{21})a_2 -(B_{12}-\widetilde{B}_{12})a_1&b_2\alpha_1+c_1\beta_2\\
A_{1\dd}^{(2)}&(-D_{21}+\widetilde{D}_{21})b_2 -(D_{12}-\widetilde{D}_{12})b_1&c_1\alpha_2+c_2\alpha_1\\
A_{1\dd}^{(3)}&(-D_{21}+\widetilde{D}_{21})c_2 -(A_{12}-\widetilde{A}_{12})c_1&\alpha_1+c_1 \rho_2\\
(\nabla_1 \Pi_1)_\dd^{(0)}&(-D_{21}+\widetilde{D}_{21})\rho_2 +(D_{21}+\widetilde{D}_{21})\rho_1 &\gamma_2 \alpha_1\\
(\nabla_1 \Pi_1)_\dd^{(1)}&(-D_{21}+\widetilde{D}_{21})\gamma_2 -(B_{12}-\widetilde{B}_{12})\gamma_1 &\beta_2\alpha_1\\
(\nabla_1 \Pi_1)_\dd^{(2)}&(-D_{21}+\widetilde{D}_{21})\beta_2 -(D_{12}-\widetilde{D}_{12})\beta_1&\alpha_1\alpha_2\\
(\nabla_1 \Pi_1)_\dd^{(3)}&(-D_{21}+\widetilde{D}_{21})\alpha_2 -(A_{12}-\widetilde{A}_{12})\alpha_1 &\alpha_1\rho_2\\
\bar{N}_2&(-D_{21} + \widetilde{D}_{21})\bar{\xi}_2 +(D_{21}+\widetilde{D}_{21})\bar{\xi}_1 &0\\
\hline
\end{array}
$$

\end{document}